\documentclass[
showpacs,superscriptaddress,twocolumn,aps,prl,longbibliography]{revtex4-1}

\usepackage[utf8]{inputenc}
\usepackage[T1]{fontenc}
\usepackage{graphicx}
\usepackage[space]{grffile}
\usepackage{dcolumn}
\usepackage{siunitx}
\DeclareSIUnit\sample{S}  
\usepackage{color}

\usepackage[hidelinks]{hyperref}
\usepackage{hyperref}

\begin{document}

\title{Single-molecule vacuum Rabi splitting:\texorpdfstring{\\}{ }four-wave mixing and optical switching at the single-photon level}

\author{André Pscherer}
\affiliation{Max Planck Institute for the Science of Light, D-91058 Erlangen, Germany}
\author{Manuel Meierhofer}
\altaffiliation[Current address: ]{Department of Physics, University of Regensburg, Universitätsstraße 31, D-93040 Regensburg, Germany}
\affiliation{Max Planck Institute for the Science of Light, D-91058 Erlangen, Germany}
\author{Daqing Wang}
\altaffiliation[Current address: ]{Experimentalphysik I, Universität Kassel, Heinrich-Plett-Straße 40, D-34132 Kassel, Germany}
\affiliation{Max Planck Institute for the Science of Light, D-91058 Erlangen, Germany}
\author{Hrishikesh Kelkar}
\altaffiliation[Current address: ]{LUMICKS~B.~V., Pilotenstraat 41, 1059CH Amsterdam, The Netherlands}
\affiliation{Max Planck Institute for the Science of Light, D-91058 Erlangen, Germany}
\author{Diego Martín-Cano}
\altaffiliation[Current address: ]{Departamento de Física Teórica de la Materia Condensada and Condensed Matter Physics Center (IFIMAC), Universidad Autónoma de Madrid, E28049 Madrid, Spain}
\affiliation{Max Planck Institute for the Science of Light, D-91058 Erlangen, Germany}
\author{Tobias Utikal}
\affiliation{Max Planck Institute for the Science of Light, D-91058 Erlangen, Germany}
\author{Stephan Götzinger}
\affiliation{Department of Physics, Friedrich-Alexander University Erlangen-Nürnberg (FAU), D-91058 Erlangen, Germany}
\affiliation{Max Planck Institute for the Science of Light, D-91058 Erlangen, Germany}
\affiliation{Graduate School in Advanced Optical Technologies (SAOT), Friedrich-Alexander University Erlangen-Nürnberg, Erlangen D-91052, Germany}
\author{Vahid Sandoghdar}
\affiliation{Max Planck Institute for the Science of Light, D-91058 Erlangen, Germany}
\affiliation{Department of Physics, Friedrich-Alexander University Erlangen-Nürnberg (FAU), D-91058 Erlangen, Germany}

\begin{abstract}

A single quantum emitter can possess a very strong intrinsic nonlinearity, but its overall promise for nonlinear effects is hampered by the challenge of efficient coupling to incident photons. Common nonlinear optical materials, on the other hand, are easy to couple to but are bulky, imposing a severe limitation on the miniaturization of photonic systems. In this work, we show that a single organic molecule acts as an extremely efficient nonlinear optical element in the strong coupling regime of cavity quantum electrodynamics. We report on single-photon sensitivity in nonlinear signal generation and all-optical switching. Our work promotes the use of molecules for applications such as integrated photonic circuits, operating at very low powers.

\end{abstract}

\maketitle

The cross section of materials for nonlinear optical processes is known to be small so that measurements are usually performed under intense laser illumination \cite{Mukamel1999book, Boyd2003}. Considering that the linear cross section of a single two-level atom ($\sigma_0=3\lambda^2/2\pi$, $\lambda$ is the transition wavelength) is large enough to result in the complete extinction of an optical beam \cite{Zumofen2008}, one might wonder about the ability of an atom or a molecule to generate nonlinear signals with single-photon sensitivity \cite{Maser2016}. It turns out, however, that $\sigma_0$ for real-life quantum emitters is compromised by the influence of many transition paths, dissipation or dephasing \cite{Loudon2000book}. To overcome the resulting decrease in coupling efficiency, single emitters such as cold alkali atoms \cite{Thompson1992}, semiconductor quantum dots \cite{Yoshie2004,Najer2019} or color centers \cite{Park2006,Janitz2020} have been investigated in the strong-coupling regime of cavity quantum electrodynamics (CQED). 

Although organic molecules were among the first nonlinear optical media that were exploited \cite{Maker1964, Schafer1966}, they have been under-represented in nonlinear CQED studies: strong coupling has been reported for ensembles of molecules \cite{Lidzey1998,Torma2015} and in one claim with single-molecule sensitivity but at a low degree of coherence \cite{Chikkaraddy2016}. In this Letter, we present the first case of strong coupling between a Fourier-limited single molecule and a microcavity. We demonstrate that this system can act as a highly efficient medium for coherent generation of nonlinear signals such as four-wave mixing and its higher harmonics as well as optical switching, at the single-photon level.

\begin{figure}[b!]
\includegraphics[width=\columnwidth]{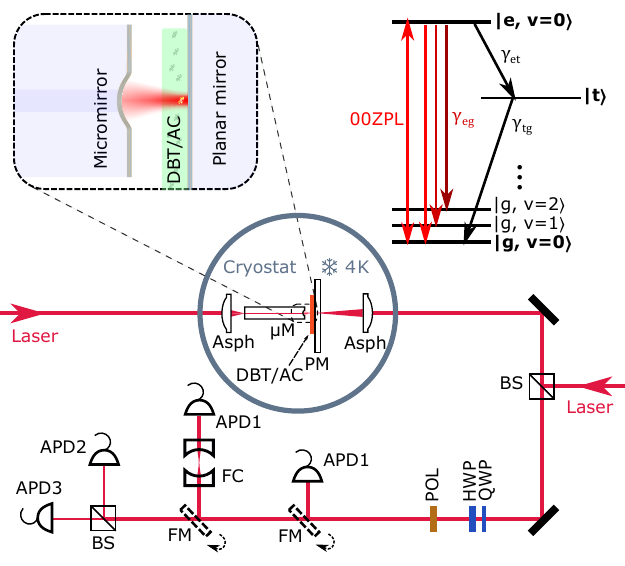}
\caption{Schematics of the experimental setup. Details can be found in Refs.\,\cite{Wang2017,Wang2019}. An external Fabry-Perot cavity (FC) serves as a narrow-band filter for the molecular emission. BS: beamsplitter, Asph: aspheric lens, µM: micromirror, PM: planar mirror, DBT/AC: dibenzoterrylene-doped anthracene crystal, QWP: quarter-wave plate, HWP: half-wave plate, POL: polarization filter, FM: flip mirror, APD: avalanche photodiode. Left inset: A close-up of the microcavity. Right inset: Jablonski diagram for DBT. The 00ZPL takes place at a wavelength of $\lambda \sim \SI{785}{\nano\meter}$. The triplet state is denoted by $|t\rangle$.}
\label{Setup}
\end{figure}

The molecule in our current work is dibenzoterrylene (DBT) from the family of polycyclic aromatic hydrocarbons (PAH). As in the case of other organic dye molecules, the excited state in DBT can decay via a manifold of vibrational levels $|g, v=0,1,2,..\rangle$ in the electronic ground state (see inset in Fig.\,\ref{Setup}(a)). When embedded in an appropriate organic crystal such as anthracene (AC), the zero-phonon line (00ZPL) associated with the transition between $|g, v=0\rangle$ and $|e, v=0\rangle$ boasts a Fourier-limited linewidth ($\gamma$) at liquid helium temperature \cite{Nicolet2007,Wang2019}. The branching ratio of this transition, defined as the ratio of the power emitted via the 00ZPL to the total fluorescence power, amounts to about \SI{30}{\percent}. To compensate for the loss of coherence through the red-shifted decay channels, we recently coupled a DBT:AC sample to a scanning Fabry-Perot microcavity and demonstrated that the composite molecule-cavity system behaves like a coherent two-level system \cite{Wang2019}. In this publication, we extend that work by entering the strong coupling regime of CQED and explore nonlinear interactions such as four-wave mixing and optical switching at the quantum level.

The experimental setup including a cryostat and various opto-electronic components was mostly as described in Refs.\,\cite{Wang2017,Wang2019}. Here, it suffices to state that we use a wavelength-sized Fabry-Perot resonator consisting of a planar mirror and a second curved mirror, which is nanofabricated at the end of an optical fiber. The mirrors surround a thin DBT:AC crystal in the cryostat (see inset in Fig.\,\ref{Setup}). To probe the transmission response of the cavity, we examined the cross-polarized signal of the light reflected from the flat mirror side or directly detected the transmission of light coupled from the fiber side \cite{Wang2017,Wang2019}. We accessed the strong coupling regime by exploiting the knowledge that the finesse of our current cavity is limited by residual mechanical instabilities (in the frequency range of \SI{10}{\hertz} - \SI{10}{\kilo\hertz}), which remain after the cavity frequency is locked using a separate laser beam \cite{Wang2017,Wang2019}. We, thus, synchronized our photon detection events with the locking error signal of the cavity to only record data in time intervals, where the cavity frequency ($\nu_{\rm c}$) coincides with the molecular 00ZPL frequency ($\nu_{\rm m}$). 

\begin{figure}[t]
\includegraphics[width=0.4\textwidth]{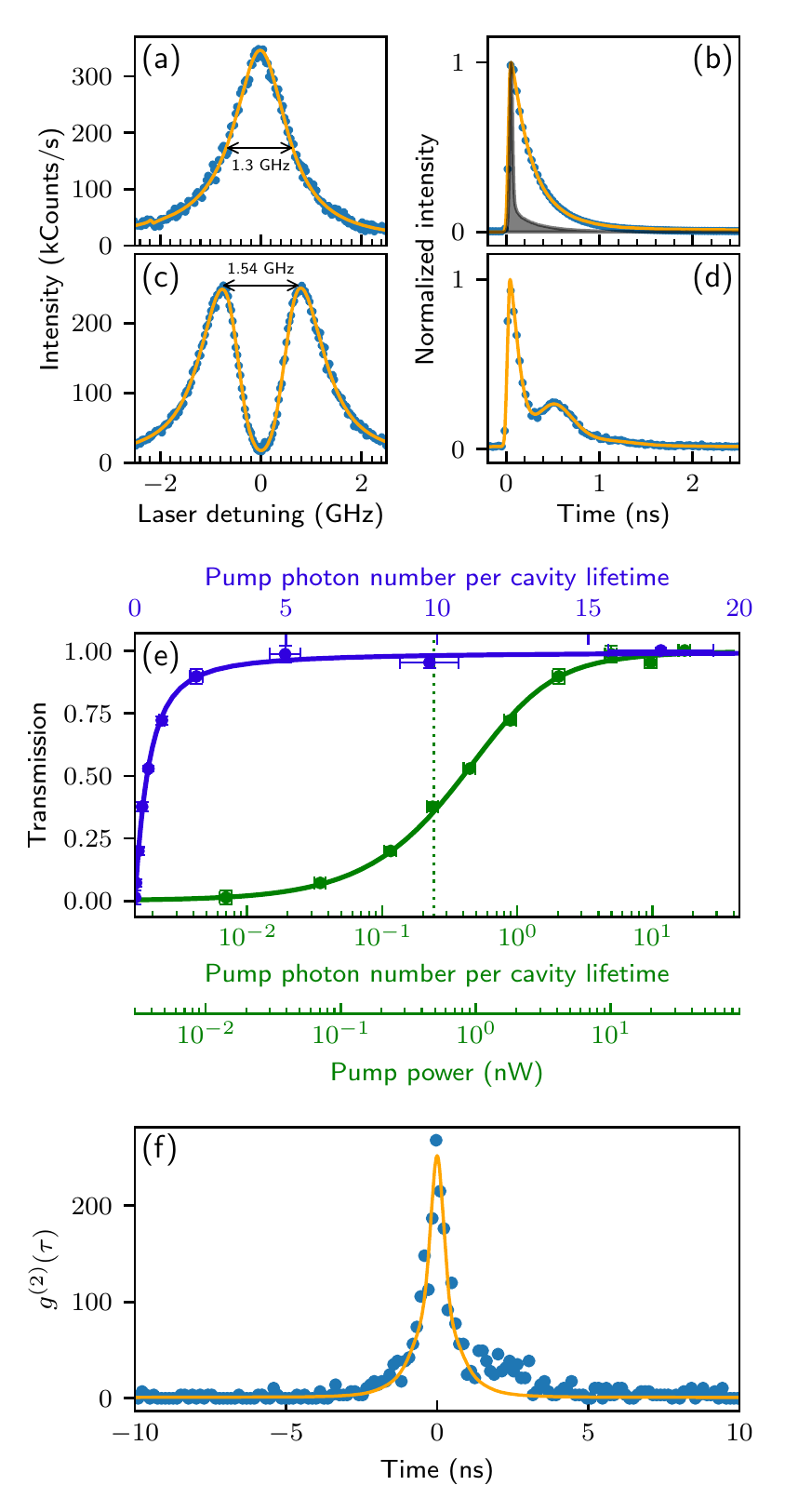}
\caption{Transmission spectrum of the bare cavity (a) and the coupled molecule-cavity system (c). Ring-down temporal signal of the bare cavity (b) and the coupled molecule-cavity system (d). The fits to the experimental data (solid orange curves) take the detector response function (shown by the grey area in (b)) into account. (e) Transmission through the cavity on resonance with the molecule and the laser as a function of the excitation power. Green and blue present the same data for different horizontal axis scalings. The vertical dotted line marks $S=1$. (f) Intensity autocorrelation of the light transmitted through the cavity on resonance with the molecule and the laser. The maximum value of 250 is limited by residual background light, which was accounted for by the theoretical fit (orange curve).} \label{StrongCoupling}
\end{figure}

Figure\,\ref{StrongCoupling}(a) displays a transmission spectrum of the microcavity as the light from a narrow-band continuous-wave titanium sapphire (Ti:Sapph) laser was coupled to the microcavity from its flat mirror end and the laser frequency was scanned. We find the full width at half-maximum (FWHM) of the cavity resonance to be $\kappa/2\pi = \SI{1.3}{\giga\hertz}$ when detuned from a molecular line. To characterize the cavity further, we also performed ring-down measurements by exciting the cavity from its optical fiber side with a picosecond pulsed Ti:Sapph laser. Figure\,\ref{StrongCoupling}(b) shows the exponential decay of the intracavity power, which yields an $e^{-1}$ decay time of \SI{125}{\pico\second} after deconvolving the instrument response function of the detector. The resulting value matches very well its Fourier correspondence given by 1/(2$\pi\times\SI{1.3}{\giga\hertz}$). We point out in passing that this decay time is considerably shorter than the excited-state lifetime of about \SI{4}{\nano\second} for DBT, i.e., $\kappa \gg \gamma \approx 2\pi \times \SI{40}{\mega\hertz} $ \cite{Wang2019}.

If we now tune the cavity resonance to the 00ZPL of a single DBT molecule, the transmission spectrum features a vacuum Rabi splitting with $2g/2\pi= \SI{1.54}{\giga\hertz}$, as displayed in Fig.\,\ref{StrongCoupling}(c). In this measurement, we accounted for a Gaussian distribution of $\SI{0.90}{\giga\hertz}$ in $\nu_{\rm c}$ caused by residual vibrations which could not be eliminated in post-processing. The corresponding time-domain measurement is shown in Fig.\,\ref{StrongCoupling}(d), where an oscillation is superimposed on the exponential cavity ring-down curve. The analysis of this signal yields a period of \SI{656}{\pico\second}, corresponding to a frequency of \SI{1.53}{\giga\hertz} and $\kappa = \SI{1.3}{\giga\hertz}$. The time-resolved oscillations provide a clear evidence that the observed splitting is indeed due to a coherent exchange of energy between the molecule and the cavity field, which is the hallmark of strong coupling in CQED \cite{Haroche2006book}. The measurements presented in Fig.\,\ref{StrongCoupling} let us extract the cooperativity parameter $C = \frac{4g^2}{\kappa \gamma} = 45$. Technical improvements in our microcavity setup will yield larger cooperativities in the near future.

Having established the regime of strong coupling, we now discuss the nonlinearity of the system. The simplest signature of nonlinearity in light-matter interactions stems from saturation, which is related to the intrinsic anharmonicity of a two-level system \cite{Boyd2003}. The green symbols in Fig.\,\ref{StrongCoupling}(e) display the cavity-molecule transmission on resonance as a function of the incident power. The solid green curve plots the results of numerical simulations based on parameters extracted from the measurements shown in Fig.\,\ref{StrongCoupling}(c),\,(d) and a careful calibration of the incident power. We reach a very good agreement with the experimental results with only the ratio of the intersystem crossing rates to ($\gamma_{\rm et}$) and out of ($\gamma_{\rm tg}$) the triplet state (see inset in Fig.\,\ref{Setup}) as a fit parameter \cite{Fleury2000,Nicolet2007}. As seen from the upper green horizontal axis in Fig.\,\ref{StrongCoupling}(e), the molecule-cavity system experiences a saturation parameter of $S=1$ for an incident average photon number as low as 0.24 per cavity lifetime. Here, we have defined $S = (\frac{\rho_{ee}(I_{\rm{in}}\to\infty)}{\rho_{ee}(I_{\rm{in}})} - 1)^{-1}$ where $\rho_{ee}$ denotes the excited state population at excitation intensity $I_{\rm{in}}$. We note that the observed behavior is similar to that of a common saturation curve from a bare molecule \cite{Wrigge2008} when plotted on a linear scale (see the blue data set in Fig.\,\ref{StrongCoupling}(e)). %$S=1$ is defined as the point, where the excited-state population reaches 1/2 of its maximum value \comment{no, 1/2 of the value for $I_{in} \to \infty$}

Another interesting consequence of the strong nonlinear response of the molecule-cavity composite is expressed by the photon statistics of the transmitted light. Figure\,\ref{StrongCoupling}(f) displays the second-order autocorrelation function $g^{(2)}(\tau)$ measured in the weak excitation regime. The observed impressive super-bunching stems from the difference in the response of the molecule-cavity system to different Fock state components $|N\rangle=|1\rangle,\,|2\rangle,\,|3\rangle$, \textit{etc.} of the laser beam \cite{Carmichael1991,Gonzalez-Tudela2015,Wang2019}. The strong response of the molecule to a single-photon state can be exploited for photon sorting \cite{Chang2007,Witthaut2012}.

\begin{figure}
\includegraphics[width=0.47\textwidth]{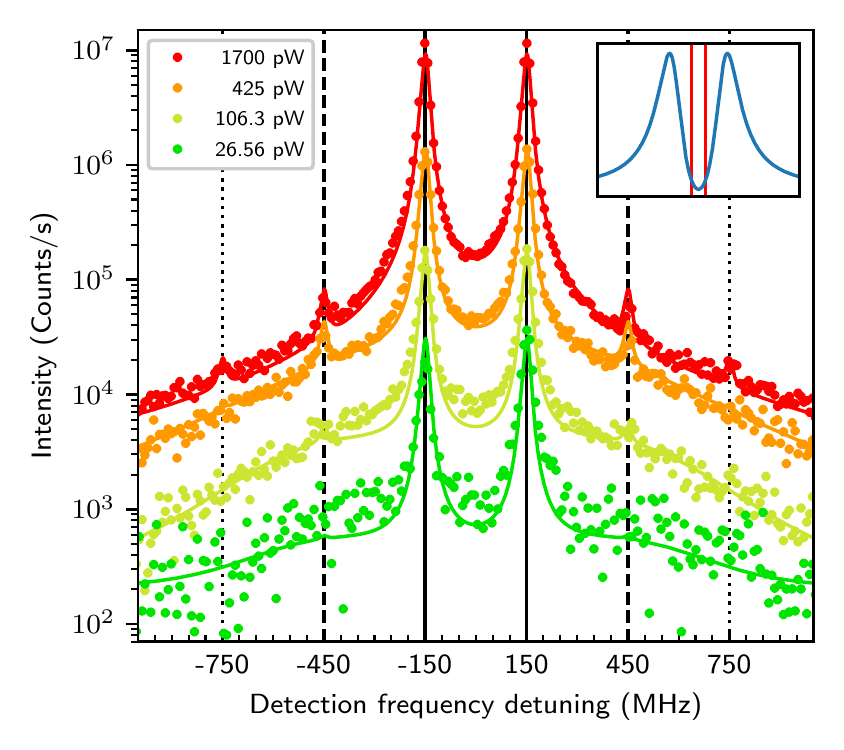}
\caption{Transmission spectrum of the coupled system under excitation at two frequencies separated by \SI{300}{\mega\hertz} at different powers (see legend) The spectra reveal the generation of higher harmonics. The theory curves reproduce the features. The vertical lines point to the regular frequency spacing of the observed signals.}\label{FWM}
\end{figure}

The results discussed in Fig.\,\ref{StrongCoupling} establish molecular CQED on the same footing as alkali atoms, semiconductor quantum dots and color center systems. We now present two investigations, where a single molecule mediates the interaction between two laser beams with average photon numbers $\overline{N}\sim 1$. First, we explore four-wave mixing (FWM) as a common nonlinear optical phenomenon. To do this, we co-coupled two laser beams at frequencies $\nu_1$ and $\nu_2$, which were detuned from $\nu_c$ by $\Delta \nu = \SI{300}{\mega\hertz}$. The powers of the two beams were equalized and their frequencies set symmetrically on each side of $\nu_c$ (see inset in Fig.\,\ref{FWM}). We then scanned an external filter cavity with a linewidth of 30\,MHz (see FC in Fig.\,\ref{Setup}) to probe the spectrum of the light exiting the molecule-cavity system. The orange data points in Fig.\,\ref{FWM} show that already at a very low cavity-coupled power of \SI{425}{\pico\watt} per beam, corresponding to 0.21 photons per cavity lifetime, we observe the conversion of a pair of photons at frequencies $(\nu_1, \nu_2)$ to a pair at frequencies $(\nu_1 - \Delta\nu, \nu_2+ \Delta\nu)$ with an efficiency ($\text{FWM peak} / \text{main peak}$) of \SI{1.4}{\percent}. This is substantially more efficient than our previous evidence of FWM produced by a molecule in a tight focus \cite{Maser2016}. The measurements in Fig.\,\ref{FWM} confirm that increasing the incident intensity lowers the efficiency since the interaction becomes less coherent for large saturation parameters \cite{Cohen-Tannoudji1998,Boyd2003,Wrigge2008}. Nevertheless, the absolute power in the FWM frequencies increases beyond $S=1$. In fact, we show that for excitation beyond saturation (\SI{1.7}{\nano\watt} per beam), a single molecule generates a six-photon process, corresponding to the detection of the second-order harmonics (see Fig.\,\ref{FWM}). In future efforts, it would be interesting to scrutinize the photons generated in the harmonics more closely to reveal their spatio-temporal entanglement \cite{Hofmann2003}. Furthermore, the efficiency of the FWM process could be enhanced by synchronizing the incident photons \cite{Dorfman2016}, e.g., through the use of triggered photon guns \cite{Chu2017}. 

\begin{figure}
\includegraphics[width=0.47\textwidth]{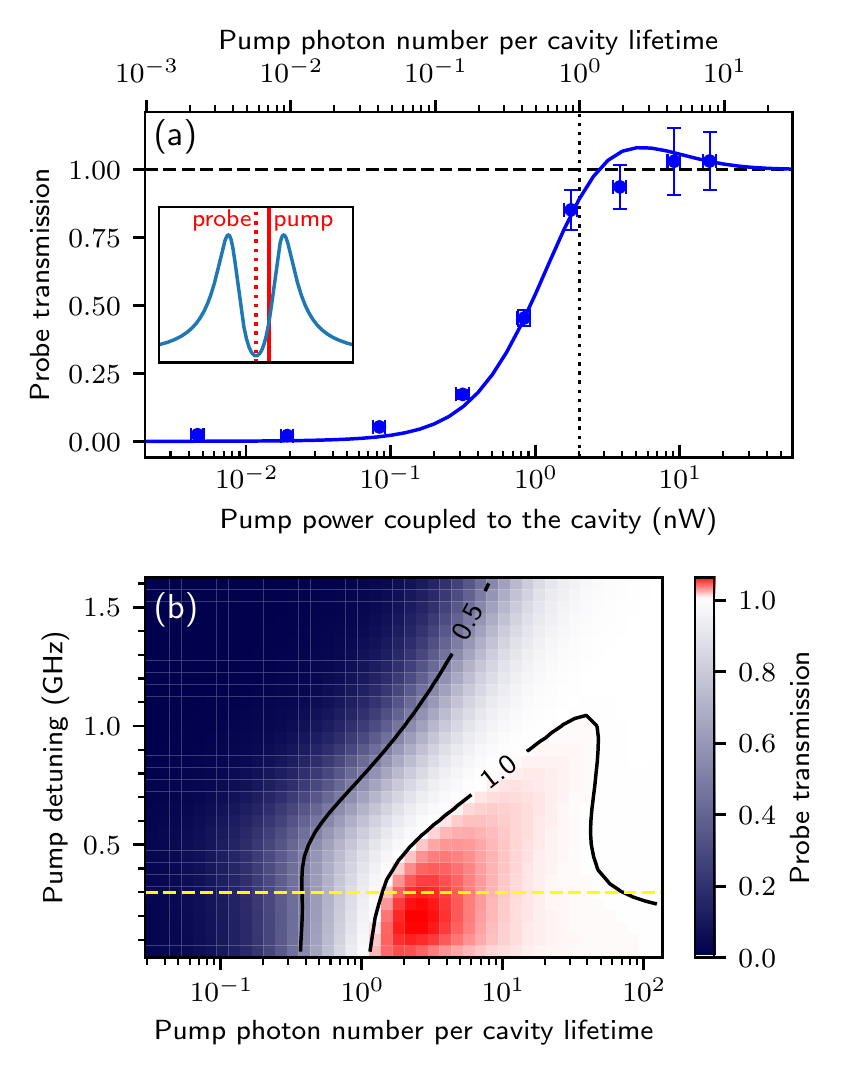}
\caption{(a) Transmission of the probe beam as a pump beam excites the molecule at a detuning of 300\,MHz. The symbols and the solid curve present the experimental and theoretical data, respectively. Measurements close to full transmission involve smaller spectral features and, thus, lead to larger error bars. (b) Calculated transmission of a probe beam as a function of the pump frequency detuning from the cavity resonance and the pump power. The horizontal dashed line indicates the conditions for measurements presented in (a).} \label{Transistor}
\end{figure}

Another powerful technique in nonlinear optics is to control the interaction between a medium and one light field (probe) via a second optical beam (pump). This approach is commonly employed in spectroscopy \cite{Mukamel1999book,Dorfman2016}, but it is also encountered in signal processing schemes, where a gate beam is used to manipulate a signal beam \cite{Gibbs1979,Hwang2009}. To explore the latter scenario, we tuned the frequency of the weaker probe beam to $\nu_{\rm c}= \nu_{\rm m}$, while the pump frequency was detuned by $\SI{300}{\mega\hertz}$ (see inset in Fig.\,\ref{Transistor}(a)). The transmission of the probe beam was then measured by scanning the filter cavity. The symbols in Fig.\,\ref{Transistor}(a) display this quantity as a function of the pump power. The solid curve presents a very good agreement between the theoretical predictions and the experimental data based on $\gamma/2\pi = \SI{0.04}{\giga\hertz}$ and independently measured parameters, $g/2\pi = \SI{0.63}{\giga\hertz}$, $\kappa/2\pi = \SI{1.3}{\giga\hertz}$, and the in-coupling efficiency of $\SI{18}{\percent}$. As in the case of Fig.\,\ref{StrongCoupling}(e), we left the ratio of the intersystem crossing rates ($\gamma_{\text{et}} / \gamma_{\text{tg}}$) as free fit parameter. Figure\,\ref{Transistor}(a) shows that with only one photon per cavity lifetime, we can nearly fully turn on the probe beam transmission, which is otherwise blocked by the molecule. We remark that transfer of energy between the two beams also allows for the amplification of the probe transmission beyond the value of 100\% \cite{Maser2016}.

The switching contrast, defined as the ratio of the transmitted powers with and without the pump, amounts to 16 dB in Fig.\,\ref{Transistor}(a), which is about 30 times higher than the best previous reports without a cavity \cite{Maser2016, Hwang2009}. To compare our study with the system response for other choices of pump and probe parameters, in Fig.\,\ref{Transistor}(b) we present the calculated value of the probe transmission that can be achieved using our system parameters at different frequency detunings and pump powers. We find that it is advantageous to choose smaller frequency detunings for achieving switching at lower power, while keeping the frequency difference large enough to be able to separate them in the detection path with high fidelity. 

Single-photon nonlinearities of quantum emitters have been considered for switching and quantum information processing \cite{Chuang1995}. Previous works have stated that perfect switching by single photons is not possible in two-level atoms due to a time-bandwidth issue and distortions of the photon wavepacket \cite{Shapiro2006,Gea-Banacloche2010} although alternative arguments have also been put forth \cite{Brod2016}. Our work inspires another intriguing approach, where single quantum emitters would be used as nano-optical logic elements for optical signal processing \cite{Hardy2007,Touch2017} with very weak light fields of average photon number $\overline{N}\sim1$. The organic solid-state platform presented in this work extends the pallet of material systems that have been used in quantum optics and offers significant advantages due to the ease of fabrication, availability in a wide range of wavelengths, brightness and Fourier-limited emission \cite{Toninelli2020arxiv}. The capacity of the organic matrices to host a very large density of dye molecules in the order of $10^4$ per \si{\micro\meter^3} makes molecular CQED easily extendable to a regime, where many emitters are strongly coupled to the same optical mode \cite{Haakh2016, Evans2018,Samutpraphoot2020}. Such an arrangement could mediate nonlinear interactions \cite{Carusotto2013} or generate $N$-photon bundles \cite{Munoz2014} in integrated photonics circuits \cite{Rattenbacher2019,Shkarin2021,Minzioni2019}. Single-molecule platforms can also be used for exploring a number of fundamental phenomena that have been predicted in the strong coupling regime of a two-level atom such as bistability \cite{SavageCarmichael1988,Rice1988}, steady-state population inversion \cite{Lindberg1988}, and single-emitter lasing \cite{Mu1992,An1994,Nomura2010,McKeever2003}. Furthermore, Fourier-limited coherence in organic molecules can be combined with plasmonic nanostructures \cite{Zirkelbach2020} and hybrid architectures \cite{Gurlek2018} to realize a nanoscopic realm of single-molecule strong coupling.

We thank Claudiu Genes and Dirk Englund for discussions, Jahangir Nobakht and Jan Renger for experimental support, and the Max Planck Society and the German Ministry of Research and Education (Grant Nr. 13N14839 within the research program ``Photonik Forschung Deutschland'') for financial support. 

%\bibliographystyle{apsrev4-1}
%\bibliography{References}

\begin{thebibliography}{75}
\makeatletter
\providecommand \@ifxundefined [1]{%
 \@ifx{#1\undefined}
}%
\providecommand \@ifnum [1]{%
 \ifnum #1\expandafter \@firstoftwo
 \else \expandafter \@secondoftwo
 \fi
}%
\providecommand \@ifx [1]{%
 \ifx #1\expandafter \@firstoftwo
 \else \expandafter \@secondoftwo
 \fi
}%
\providecommand \natexlab [1]{#1}%
\providecommand \enquote  [1]{``#1''}%
\providecommand \bibnamefont  [1]{#1}%
\providecommand \bibfnamefont [1]{#1}%
\providecommand \citenamefont [1]{#1}%
\providecommand \href@noop [0]{\@secondoftwo}%
\providecommand \href [0]{\begingroup \@sanitize@url \@href}%
\providecommand \@href[1]{\@@startlink{#1}\@@href}%
\providecommand \@@href[1]{\endgroup#1\@@endlink}%
\providecommand \@sanitize@url [0]{\catcode `\\12\catcode `\$12\catcode
  `\&12\catcode `\#12\catcode `\^12\catcode `\_12\catcode `\%12\relax}%
\providecommand \@@startlink[1]{}%
\providecommand \@@endlink[0]{}%
\providecommand \url  [0]{\begingroup\@sanitize@url \@url }%
\providecommand \@url [1]{\endgroup\@href {#1}{\urlprefix }}%
\providecommand \urlprefix  [0]{URL }%
\providecommand \Eprint [0]{\href }%
\providecommand \doibase [0]{http://dx.doi.org/}%
\providecommand \selectlanguage [0]{\@gobble}%
\providecommand \bibinfo  [0]{\@secondoftwo}%
\providecommand \bibfield  [0]{\@secondoftwo}%
\providecommand \translation [1]{[#1]}%
\providecommand \BibitemOpen [0]{}%
\providecommand \bibitemStop [0]{}%
\providecommand \bibitemNoStop [0]{.\EOS\space}%
\providecommand \EOS [0]{\spacefactor3000\relax}%
\providecommand \BibitemShut  [1]{\csname bibitem#1\endcsname}%
\let\auto@bib@innerbib\@empty
%</preamble>
\bibitem [{\citenamefont {Mukamel}(1999)}]{Mukamel1999book}%
  \BibitemOpen
  \bibfield  {author} {\bibinfo {author} {\bibfnamefont {S.}~\bibnamefont
  {Mukamel}},\ }\href@noop {} {\emph {\bibinfo {title} {Principles of Nonlinear
  Optical Spectroscopy}}},\ Oxford series in optical and imaging sciences\
  (\bibinfo  {publisher} {Oxford University Press},\ \bibinfo {year}
  {1999})\BibitemShut {NoStop}%
\bibitem [{\citenamefont {Boyd}(2003)}]{Boyd2003}%
  \BibitemOpen
  \bibfield  {author} {\bibinfo {author} {\bibfnamefont {R.~W.}\ \bibnamefont
  {Boyd}},\ }\href@noop {} {\emph {\bibinfo {title} {{Nonlinear optics}}}}\
  (\bibinfo  {publisher} {Elsevier},\ \bibinfo {year} {2003})\BibitemShut
  {NoStop}%
\bibitem [{\citenamefont {Zumofen}\ \emph {et~al.}(2008)\citenamefont
  {Zumofen}, \citenamefont {Mojarad}, \citenamefont {Sandoghdar},\ and\
  \citenamefont {Agio}}]{Zumofen2008}%
  \BibitemOpen
  \bibfield  {author} {\bibinfo {author} {\bibfnamefont {G.}~\bibnamefont
  {Zumofen}}, \bibinfo {author} {\bibfnamefont {N.~M.}\ \bibnamefont
  {Mojarad}}, \bibinfo {author} {\bibfnamefont {V.}~\bibnamefont {Sandoghdar}},
  \ and\ \bibinfo {author} {\bibfnamefont {M}~\bibnamefont {Agio}},\ }\bibfield
   {title} {\enquote {\bibinfo {title} {{Perfect Reflection of Light by an
  Oscillating Dipole}},}\ }\href {\doibase 10.1103/PhysRevLett.101.180404}
  {\bibfield  {journal} {\bibinfo  {journal} {Phys. Rev. Lett.}\ }\textbf
  {\bibinfo {volume} {101}},\ \bibinfo {pages} {180404} (\bibinfo {year}
  {2008})}\BibitemShut {NoStop}%
\bibitem [{\citenamefont {Maser}\ \emph {et~al.}(2016)\citenamefont {Maser},
  \citenamefont {Gmeiner}, \citenamefont {Utikal}, \citenamefont {Götzinger},\
  and\ \citenamefont {Sandoghdar}}]{Maser2016}%
  \BibitemOpen
  \bibfield  {author} {\bibinfo {author} {\bibfnamefont {A.}~\bibnamefont
  {Maser}}, \bibinfo {author} {\bibfnamefont {B.}~\bibnamefont {Gmeiner}},
  \bibinfo {author} {\bibfnamefont {T.}~\bibnamefont {Utikal}}, \bibinfo
  {author} {\bibfnamefont {S.}~\bibnamefont {Götzinger}}, \ and\ \bibinfo
  {author} {\bibfnamefont {V.}~\bibnamefont {Sandoghdar}},\ }\bibfield  {title}
  {\enquote {\bibinfo {title} {{Few-photon coherent nonlinear optics with a
  single molecule}},}\ }\href {\doibase 10.1038/nphoton.2016.63} {\bibfield
  {journal} {\bibinfo  {journal} {Nat. Photonics}\ }\textbf {\bibinfo {volume}
  {10}},\ \bibinfo {pages} {450--453} (\bibinfo {year} {2016})}\BibitemShut
  {NoStop}%
\bibitem [{\citenamefont {Loudon}(2000)}]{Loudon2000book}%
  \BibitemOpen
  \bibfield  {author} {\bibinfo {author} {\bibfnamefont {R.}~\bibnamefont
  {Loudon}},\ }\href@noop {} {\emph {\bibinfo {title} {{The Quantum Theory of
  Light}}}},\ \bibinfo {edition} {3rd}\ ed.\ (\bibinfo  {publisher} {Oxford
  University Press},\ \bibinfo {address} {New York},\ \bibinfo {year}
  {2000})\BibitemShut {NoStop}%
\bibitem [{\citenamefont {Thompson}\ \emph {et~al.}(1992)\citenamefont
  {Thompson}, \citenamefont {Rempe},\ and\ \citenamefont
  {Kimble}}]{Thompson1992}%
  \BibitemOpen
  \bibfield  {author} {\bibinfo {author} {\bibfnamefont {R.~J.}\ \bibnamefont
  {Thompson}}, \bibinfo {author} {\bibfnamefont {G.}~\bibnamefont {Rempe}}, \
  and\ \bibinfo {author} {\bibfnamefont {H.~J.}\ \bibnamefont {Kimble}},\
  }\bibfield  {title} {\enquote {\bibinfo {title} {{Observation of normal-mode
  splitting for an atom in an optical cavity}},}\ }\href {\doibase
  10.1103/PhysRevLett.68.1132} {\bibfield  {journal} {\bibinfo  {journal}
  {Phys. Rev. Lett.}\ }\textbf {\bibinfo {volume} {68}},\ \bibinfo {pages}
  {1132--1135} (\bibinfo {year} {1992})}\BibitemShut {NoStop}%
\bibitem [{\citenamefont {Yoshie}\ \emph {et~al.}(2004)\citenamefont {Yoshie},
  \citenamefont {Scherer}, \citenamefont {Hendrickson}, \citenamefont
  {Khitrova}, \citenamefont {Gibbs}, \citenamefont {Rupper}, \citenamefont
  {Ell}, \citenamefont {Shchekin},\ and\ \citenamefont {Deppe}}]{Yoshie2004}%
  \BibitemOpen
  \bibfield  {author} {\bibinfo {author} {\bibfnamefont {T.}~\bibnamefont
  {Yoshie}}, \bibinfo {author} {\bibfnamefont {A.}~\bibnamefont {Scherer}},
  \bibinfo {author} {\bibfnamefont {J.}~\bibnamefont {Hendrickson}}, \bibinfo
  {author} {\bibfnamefont {G.}~\bibnamefont {Khitrova}}, \bibinfo {author}
  {\bibfnamefont {H.~M.}\ \bibnamefont {Gibbs}}, \bibinfo {author}
  {\bibfnamefont {G.}~\bibnamefont {Rupper}}, \bibinfo {author} {\bibfnamefont
  {C.}~\bibnamefont {Ell}}, \bibinfo {author} {\bibfnamefont {O.~B.}\
  \bibnamefont {Shchekin}}, \ and\ \bibinfo {author} {\bibfnamefont {D.~G.}\
  \bibnamefont {Deppe}},\ }\bibfield  {title} {\enquote {\bibinfo {title}
  {{Vacuum Rabi splitting with a single quantum dot in a photonic crystal
  nanocavity}},}\ }\href {\doibase 10.1038/nature03119} {\bibfield  {journal}
  {\bibinfo  {journal} {Nature}\ }\textbf {\bibinfo {volume} {432}},\ \bibinfo
  {pages} {200--203} (\bibinfo {year} {2004})}\BibitemShut {NoStop}%
\bibitem [{\citenamefont {Najer}\ \emph {et~al.}(2019)\citenamefont {Najer},
  \citenamefont {Söllner}, \citenamefont {Sekatski}, \citenamefont {Dolique},
  \citenamefont {Löbl}, \citenamefont {Riedel}, \citenamefont {Schott},
  \citenamefont {Starosielec}, \citenamefont {Valentin}, \citenamefont {Wieck},
  \citenamefont {Sangouard}, \citenamefont {Ludwig},\ and\ \citenamefont
  {Warburton}}]{Najer2019}%
  \BibitemOpen
  \bibfield  {author} {\bibinfo {author} {\bibfnamefont {D.}~\bibnamefont
  {Najer}}, \bibinfo {author} {\bibfnamefont {I.}~\bibnamefont {Söllner}},
  \bibinfo {author} {\bibfnamefont {P.}~\bibnamefont {Sekatski}}, \bibinfo
  {author} {\bibfnamefont {V.}~\bibnamefont {Dolique}}, \bibinfo {author}
  {\bibfnamefont {M.~C.}\ \bibnamefont {Löbl}}, \bibinfo {author}
  {\bibfnamefont {D.}~\bibnamefont {Riedel}}, \bibinfo {author} {\bibfnamefont
  {R.}~\bibnamefont {Schott}}, \bibinfo {author} {\bibfnamefont
  {S.}~\bibnamefont {Starosielec}}, \bibinfo {author} {\bibfnamefont {S.~R.}\
  \bibnamefont {Valentin}}, \bibinfo {author} {\bibfnamefont {A.~D.}\
  \bibnamefont {Wieck}}, \bibinfo {author} {\bibfnamefont {N.}~\bibnamefont
  {Sangouard}}, \bibinfo {author} {\bibfnamefont {A.}~\bibnamefont {Ludwig}}, \
  and\ \bibinfo {author} {\bibfnamefont {R.~J.}\ \bibnamefont {Warburton}},\
  }\bibfield  {title} {\enquote {\bibinfo {title} {{A gated quantum dot
  strongly coupled to an optical microcavity}},}\ }\href {\doibase
  10.1038/s41586-019-1709-y} {\bibfield  {journal} {\bibinfo  {journal}
  {Nature}\ }\textbf {\bibinfo {volume} {575}},\ \bibinfo {pages} {622--627}
  (\bibinfo {year} {2019})}\BibitemShut {NoStop}%
\bibitem [{\citenamefont {Park}\ \emph {et~al.}(2006)\citenamefont {Park},
  \citenamefont {Cook},\ and\ \citenamefont {Wang}}]{Park2006}%
  \BibitemOpen
  \bibfield  {author} {\bibinfo {author} {\bibfnamefont {Y.-S.}\ \bibnamefont
  {Park}}, \bibinfo {author} {\bibfnamefont {A.~K.}\ \bibnamefont {Cook}}, \
  and\ \bibinfo {author} {\bibfnamefont {H.}~\bibnamefont {Wang}},\ }\bibfield
  {title} {\enquote {\bibinfo {title} {{Cavity QED with Diamond Nanocrystals
  and Silica Microspheres}},}\ }\href {\doibase 10.1021/nl061342r} {\bibfield
  {journal} {\bibinfo  {journal} {Nano Lett.}\ }\textbf {\bibinfo {volume}
  {6}},\ \bibinfo {pages} {2075--2079} (\bibinfo {year} {2006})}\BibitemShut
  {NoStop}%
\bibitem [{\citenamefont {Janitz}\ \emph {et~al.}(2020)\citenamefont {Janitz},
  \citenamefont {Bhaskar},\ and\ \citenamefont {Childress}}]{Janitz2020}%
  \BibitemOpen
  \bibfield  {author} {\bibinfo {author} {\bibfnamefont {E.}~\bibnamefont
  {Janitz}}, \bibinfo {author} {\bibfnamefont {M.~K.}\ \bibnamefont {Bhaskar}},
  \ and\ \bibinfo {author} {\bibfnamefont {L.}~\bibnamefont {Childress}},\
  }\bibfield  {title} {\enquote {\bibinfo {title} {{Cavity quantum
  electrodynamics with color centers in diamond}},}\ }\href {\doibase
  10.1364/OPTICA.398628} {\bibfield  {journal} {\bibinfo  {journal} {Optica}\
  }\textbf {\bibinfo {volume} {7}},\ \bibinfo {pages} {1232} (\bibinfo {year}
  {2020})}\BibitemShut {NoStop}%
\bibitem [{\citenamefont {Maker}\ \emph {et~al.}(1964)\citenamefont {Maker},
  \citenamefont {Terhune},\ and\ \citenamefont {Savage}}]{Maker1964}%
  \BibitemOpen
  \bibfield  {author} {\bibinfo {author} {\bibfnamefont {P.~D.}\ \bibnamefont
  {Maker}}, \bibinfo {author} {\bibfnamefont {R.~W.}\ \bibnamefont {Terhune}},
  \ and\ \bibinfo {author} {\bibfnamefont {C.~M.}\ \bibnamefont {Savage}},\
  }\bibfield  {title} {\enquote {\bibinfo {title} {{Intensity-Dependent Changes
  in the Refractive Index of Liquids}},}\ }\href {\doibase
  10.1103/PhysRevLett.12.507} {\bibfield  {journal} {\bibinfo  {journal} {Phys.
  Rev. Lett.}\ }\textbf {\bibinfo {volume} {12}},\ \bibinfo {pages} {507--509}
  (\bibinfo {year} {1964})}\BibitemShut {NoStop}%
\bibitem [{\citenamefont {Schäfer}\ \emph {et~al.}(1966)\citenamefont
  {Schäfer}, \citenamefont {Schmidt},\ and\ \citenamefont
  {Volze}}]{Schafer1966}%
  \BibitemOpen
  \bibfield  {author} {\bibinfo {author} {\bibfnamefont {F.~P.}\ \bibnamefont
  {Schäfer}}, \bibinfo {author} {\bibfnamefont {W.}~\bibnamefont {Schmidt}}, \
  and\ \bibinfo {author} {\bibfnamefont {J.}~\bibnamefont {Volze}},\ }\bibfield
   {title} {\enquote {\bibinfo {title} {{Organic Dye Solution Laser}},}\ }\href
  {\doibase 10.1063/1.1754762} {\bibfield  {journal} {\bibinfo  {journal}
  {Appl. Phys. Lett.}\ }\textbf {\bibinfo {volume} {9}},\ \bibinfo {pages}
  {306--309} (\bibinfo {year} {1966})}\BibitemShut {NoStop}%
\bibitem [{\citenamefont {Lidzey}\ \emph {et~al.}(1998)\citenamefont {Lidzey},
  \citenamefont {Bradley}, \citenamefont {Skolnick}, \citenamefont {Virgili},
  \citenamefont {Walker},\ and\ \citenamefont {Whittaker}}]{Lidzey1998}%
  \BibitemOpen
  \bibfield  {author} {\bibinfo {author} {\bibfnamefont {D.~G.}\ \bibnamefont
  {Lidzey}}, \bibinfo {author} {\bibfnamefont {D.~D.~C.}\ \bibnamefont
  {Bradley}}, \bibinfo {author} {\bibfnamefont {M.~S.}\ \bibnamefont
  {Skolnick}}, \bibinfo {author} {\bibfnamefont {T.}~\bibnamefont {Virgili}},
  \bibinfo {author} {\bibfnamefont {S.}~\bibnamefont {Walker}}, \ and\ \bibinfo
  {author} {\bibfnamefont {D.~M.}\ \bibnamefont {Whittaker}},\ }\bibfield
  {title} {\enquote {\bibinfo {title} {{Strong exciton–photon coupling in an
  organic semiconductor microcavity}},}\ }\href {\doibase 10.1038/25692}
  {\bibfield  {journal} {\bibinfo  {journal} {Nature}\ }\textbf {\bibinfo
  {volume} {395}},\ \bibinfo {pages} {53--55} (\bibinfo {year}
  {1998})}\BibitemShut {NoStop}%
\bibitem [{\citenamefont {Törmä}\ and\ \citenamefont
  {Barnes}(2015)}]{Torma2015}%
  \BibitemOpen
  \bibfield  {author} {\bibinfo {author} {\bibfnamefont {P.}~\bibnamefont
  {Törmä}}\ and\ \bibinfo {author} {\bibfnamefont {W.~L.}\ \bibnamefont
  {Barnes}},\ }\bibfield  {title} {\enquote {\bibinfo {title} {{Strong coupling
  between surface plasmon polaritons and emitters: a review}},}\ }\href
  {\doibase 10.1088/0034-4885/78/1/013901} {\bibfield  {journal} {\bibinfo
  {journal} {Rep. Prog. Phys.}\ }\textbf {\bibinfo {volume} {78}},\ \bibinfo
  {pages} {013901} (\bibinfo {year} {2015})}\BibitemShut {NoStop}%
\bibitem [{\citenamefont {Chikkaraddy}\ \emph {et~al.}(2016)\citenamefont
  {Chikkaraddy}, \citenamefont {de~Nijs}, \citenamefont {Benz}, \citenamefont
  {Barrow}, \citenamefont {Scherman}, \citenamefont {Rosta}, \citenamefont
  {Demetriadou}, \citenamefont {Fox}, \citenamefont {Hess},\ and\ \citenamefont
  {Baumberg}}]{Chikkaraddy2016}%
  \BibitemOpen
  \bibfield  {author} {\bibinfo {author} {\bibfnamefont {R.}~\bibnamefont
  {Chikkaraddy}}, \bibinfo {author} {\bibfnamefont {B.}~\bibnamefont
  {de~Nijs}}, \bibinfo {author} {\bibfnamefont {F.}~\bibnamefont {Benz}},
  \bibinfo {author} {\bibfnamefont {S.~J.}\ \bibnamefont {Barrow}}, \bibinfo
  {author} {\bibfnamefont {O.~A.}\ \bibnamefont {Scherman}}, \bibinfo {author}
  {\bibfnamefont {E.}~\bibnamefont {Rosta}}, \bibinfo {author} {\bibfnamefont
  {A.}~\bibnamefont {Demetriadou}}, \bibinfo {author} {\bibfnamefont
  {P.}~\bibnamefont {Fox}}, \bibinfo {author} {\bibfnamefont {O.}~\bibnamefont
  {Hess}}, \ and\ \bibinfo {author} {\bibfnamefont {J.~J.}\ \bibnamefont
  {Baumberg}},\ }\bibfield  {title} {\enquote {\bibinfo {title}
  {{Single-molecule strong coupling at room temperature in plasmonic
  nanocavities}},}\ }\href {\doibase 10.1038/nature17974} {\bibfield  {journal}
  {\bibinfo  {journal} {Nature}\ }\textbf {\bibinfo {volume} {535}},\ \bibinfo
  {pages} {127--130} (\bibinfo {year} {2016})}\BibitemShut {NoStop}%
\bibitem [{\citenamefont {Wang}\ \emph {et~al.}(2017)\citenamefont {Wang},
  \citenamefont {Kelkar}, \citenamefont {Martín-Cano}, \citenamefont {Utikal},
  \citenamefont {Götzinger},\ and\ \citenamefont {Sandoghdar}}]{Wang2017}%
  \BibitemOpen
  \bibfield  {author} {\bibinfo {author} {\bibfnamefont {D.}~\bibnamefont
  {Wang}}, \bibinfo {author} {\bibfnamefont {H.}~\bibnamefont {Kelkar}},
  \bibinfo {author} {\bibfnamefont {D.}~\bibnamefont {Martín-Cano}}, \bibinfo
  {author} {\bibfnamefont {T.}~\bibnamefont {Utikal}}, \bibinfo {author}
  {\bibfnamefont {S.}~\bibnamefont {Götzinger}}, \ and\ \bibinfo {author}
  {\bibfnamefont {V.}~\bibnamefont {Sandoghdar}},\ }\bibfield  {title}
  {\enquote {\bibinfo {title} {{Coherent Coupling of a Single Molecule to a
  Scanning Fabry-Perot Microcavity}},}\ }\href {\doibase
  10.1103/PhysRevX.7.021014} {\bibfield  {journal} {\bibinfo  {journal} {Phys.
  Rev. X}\ }\textbf {\bibinfo {volume} {7}},\ \bibinfo {pages} {021014}
  (\bibinfo {year} {2017})}\BibitemShut {NoStop}%
\bibitem [{\citenamefont {Wang}\ \emph {et~al.}(2019)\citenamefont {Wang},
  \citenamefont {Kelkar}, \citenamefont {Martín-Cano}, \citenamefont
  {Rattenbacher}, \citenamefont {Shkarin}, \citenamefont {Utikal},
  \citenamefont {Götzinger},\ and\ \citenamefont {Sandoghdar}}]{Wang2019}%
  \BibitemOpen
  \bibfield  {author} {\bibinfo {author} {\bibfnamefont {D.}~\bibnamefont
  {Wang}}, \bibinfo {author} {\bibfnamefont {H.}~\bibnamefont {Kelkar}},
  \bibinfo {author} {\bibfnamefont {D.}~\bibnamefont {Martín-Cano}}, \bibinfo
  {author} {\bibfnamefont {D.}~\bibnamefont {Rattenbacher}}, \bibinfo {author}
  {\bibfnamefont {A.}~\bibnamefont {Shkarin}}, \bibinfo {author} {\bibfnamefont
  {T.}~\bibnamefont {Utikal}}, \bibinfo {author} {\bibfnamefont
  {S.}~\bibnamefont {Götzinger}}, \ and\ \bibinfo {author} {\bibfnamefont
  {V.}~\bibnamefont {Sandoghdar}},\ }\bibfield  {title} {\enquote {\bibinfo
  {title} {{Turning a molecule into a coherent two-level quantum system}},}\
  }\href {\doibase 10.1038/s41567-019-0436-5} {\bibfield  {journal} {\bibinfo
  {journal} {Nat. Phys.}\ }\textbf {\bibinfo {volume} {15}},\ \bibinfo {pages}
  {483--489} (\bibinfo {year} {2019})}\BibitemShut {NoStop}%
\bibitem [{\citenamefont {Nicolet}\ \emph {et~al.}(2007)\citenamefont
  {Nicolet}, \citenamefont {Hofmann}, \citenamefont {Kol'chenko}, \citenamefont
  {Kozankiewicz},\ and\ \citenamefont {Orrit}}]{Nicolet2007}%
  \BibitemOpen
  \bibfield  {author} {\bibinfo {author} {\bibfnamefont {A.~A.~L.}\
  \bibnamefont {Nicolet}}, \bibinfo {author} {\bibfnamefont {C.}~\bibnamefont
  {Hofmann}}, \bibinfo {author} {\bibfnamefont {M.~A.}\ \bibnamefont
  {Kol'chenko}}, \bibinfo {author} {\bibfnamefont {B.}~\bibnamefont
  {Kozankiewicz}}, \ and\ \bibinfo {author} {\bibfnamefont {M.}~\bibnamefont
  {Orrit}},\ }\bibfield  {title} {\enquote {\bibinfo {title} {{Single
  dibenzoterrylene molecules in an anthracene crystal: Spectroscopy and
  photophysics}},}\ }\href {\doibase 10.1002/cphc.200700091} {\bibfield
  {journal} {\bibinfo  {journal} {ChemPhysChem}\ }\textbf {\bibinfo {volume}
  {8}},\ \bibinfo {pages} {1215--1220} (\bibinfo {year} {2007})}\BibitemShut
  {NoStop}%
\bibitem [{\citenamefont {Haroche}\ and\ \citenamefont
  {Raimond}(2006)}]{Haroche2006book}%
  \BibitemOpen
  \bibfield  {author} {\bibinfo {author} {\bibfnamefont {S.}~\bibnamefont
  {Haroche}}\ and\ \bibinfo {author} {\bibfnamefont {J.-M.}\ \bibnamefont
  {Raimond}},\ }\href@noop {} {\emph {\bibinfo {title} {{Exploring the Quantum:
  Atoms, Cavities, and Photons}}}}\ (\bibinfo  {publisher} {Oxford University
  Press},\ \bibinfo {year} {2006})\BibitemShut {NoStop}%
\bibitem [{\citenamefont {Fleury}\ \emph {et~al.}(2000)\citenamefont {Fleury},
  \citenamefont {Segura}, \citenamefont {Zumofen}, \citenamefont {Hecht},\ and\
  \citenamefont {Wild}}]{Fleury2000}%
  \BibitemOpen
  \bibfield  {author} {\bibinfo {author} {\bibfnamefont {L.}~\bibnamefont
  {Fleury}}, \bibinfo {author} {\bibfnamefont {J.-M.}\ \bibnamefont {Segura}},
  \bibinfo {author} {\bibfnamefont {G.}~\bibnamefont {Zumofen}}, \bibinfo
  {author} {\bibfnamefont {B.}~\bibnamefont {Hecht}}, \ and\ \bibinfo {author}
  {\bibfnamefont {U.~P.}\ \bibnamefont {Wild}},\ }\bibfield  {title} {\enquote
  {\bibinfo {title} {{Nonclassical Photon Statistics in Single-Molecule
  Fluorescence at Room Temperature}},}\ }\href {\doibase
  10.1103/PhysRevLett.84.1148} {\bibfield  {journal} {\bibinfo  {journal}
  {Phys. Rev. Lett.}\ }\textbf {\bibinfo {volume} {84}},\ \bibinfo {pages}
  {1148--1151} (\bibinfo {year} {2000})}\BibitemShut {NoStop}%
\bibitem [{\citenamefont {Wrigge}\ \emph {et~al.}(2008)\citenamefont {Wrigge},
  \citenamefont {Gerhardt}, \citenamefont {Hwang}, \citenamefont {Zumofen},\
  and\ \citenamefont {Sandoghdar}}]{Wrigge2008}%
  \BibitemOpen
  \bibfield  {author} {\bibinfo {author} {\bibfnamefont {G.}~\bibnamefont
  {Wrigge}}, \bibinfo {author} {\bibfnamefont {I.}~\bibnamefont {Gerhardt}},
  \bibinfo {author} {\bibfnamefont {J.}~\bibnamefont {Hwang}}, \bibinfo
  {author} {\bibfnamefont {G.}~\bibnamefont {Zumofen}}, \ and\ \bibinfo
  {author} {\bibfnamefont {V.}~\bibnamefont {Sandoghdar}},\ }\bibfield  {title}
  {\enquote {\bibinfo {title} {{Efficient coupling of photons to a single
  molecule and the observation of its resonance fluorescence}},}\ }\href
  {\doibase 10.1038/nphys812} {\bibfield  {journal} {\bibinfo  {journal} {Nat.
  Phys.}\ }\textbf {\bibinfo {volume} {4}},\ \bibinfo {pages} {60--66}
  (\bibinfo {year} {2008})}\BibitemShut {NoStop}%
\bibitem [{\citenamefont {Carmichael}\ \emph {et~al.}(1991)\citenamefont
  {Carmichael}, \citenamefont {Brecha},\ and\ \citenamefont
  {Rice}}]{Carmichael1991}%
  \BibitemOpen
  \bibfield  {author} {\bibinfo {author} {\bibfnamefont {H.~J.}\ \bibnamefont
  {Carmichael}}, \bibinfo {author} {\bibfnamefont {R.~J.}\ \bibnamefont
  {Brecha}}, \ and\ \bibinfo {author} {\bibfnamefont {P.~R.}\ \bibnamefont
  {Rice}},\ }\bibfield  {title} {\enquote {\bibinfo {title} {{Quantum
  interference and collapse of the wavefunction in cavity QED}},}\ }\href
  {\doibase 10.1016/0030-4018(91)90194-I} {\bibfield  {journal} {\bibinfo
  {journal} {Opt. Commun.}\ }\textbf {\bibinfo {volume} {82}},\ \bibinfo
  {pages} {73--79} (\bibinfo {year} {1991})}\BibitemShut {NoStop}%
\bibitem [{\citenamefont {González-Tudela}\ \emph {et~al.}(2015)\citenamefont
  {González-Tudela}, \citenamefont {del Valle},\ and\ \citenamefont
  {Laussy}}]{Gonzalez-Tudela2015}%
  \BibitemOpen
  \bibfield  {author} {\bibinfo {author} {\bibfnamefont {A.}~\bibnamefont
  {González-Tudela}}, \bibinfo {author} {\bibfnamefont {E.}~\bibnamefont {del
  Valle}}, \ and\ \bibinfo {author} {\bibfnamefont {F.~P.}\ \bibnamefont
  {Laussy}},\ }\bibfield  {title} {\enquote {\bibinfo {title} {{Optimization of
  photon correlations by frequency filtering}},}\ }\href {\doibase
  10.1103/PhysRevA.91.043807} {\bibfield  {journal} {\bibinfo  {journal} {Phys.
  Rev. A}\ }\textbf {\bibinfo {volume} {91}},\ \bibinfo {pages} {043807}
  (\bibinfo {year} {2015})}\BibitemShut {NoStop}%
\bibitem [{\citenamefont {Chang}\ \emph {et~al.}(2007)\citenamefont {Chang},
  \citenamefont {Sørensen}, \citenamefont {Demler},\ and\ \citenamefont
  {Lukin}}]{Chang2007}%
  \BibitemOpen
  \bibfield  {author} {\bibinfo {author} {\bibfnamefont {D.~E.}\ \bibnamefont
  {Chang}}, \bibinfo {author} {\bibfnamefont {A.~S.}\ \bibnamefont
  {Sørensen}}, \bibinfo {author} {\bibfnamefont {E.~A.}\ \bibnamefont
  {Demler}}, \ and\ \bibinfo {author} {\bibfnamefont {M.~D.}\ \bibnamefont
  {Lukin}},\ }\bibfield  {title} {\enquote {\bibinfo {title} {{A single-photon
  transistor using nanoscale surface plasmons}},}\ }\href {\doibase
  10.1038/nphys708} {\bibfield  {journal} {\bibinfo  {journal} {Nat. Phys.}\
  }\textbf {\bibinfo {volume} {3}},\ \bibinfo {pages} {807--812} (\bibinfo
  {year} {2007})}\BibitemShut {NoStop}%
\bibitem [{\citenamefont {Witthaut}\ \emph {et~al.}(2012)\citenamefont
  {Witthaut}, \citenamefont {Lukin},\ and\ \citenamefont
  {Sørensen}}]{Witthaut2012}%
  \BibitemOpen
  \bibfield  {author} {\bibinfo {author} {\bibfnamefont {D.}~\bibnamefont
  {Witthaut}}, \bibinfo {author} {\bibfnamefont {M.~D.}\ \bibnamefont {Lukin}},
  \ and\ \bibinfo {author} {\bibfnamefont {A.~S.}\ \bibnamefont {Sørensen}},\
  }\bibfield  {title} {\enquote {\bibinfo {title} {{Photon sorters and QND
  detectors using single photon emitters}},}\ }\href {\doibase
  10.1209/0295-5075/97/50007} {\bibfield  {journal} {\bibinfo  {journal}
  {Europhys. Lett.}\ }\textbf {\bibinfo {volume} {97}},\ \bibinfo {pages}
  {50007} (\bibinfo {year} {2012})}\BibitemShut {NoStop}%
\bibitem [{\citenamefont {Cohen-Tannoudji}\ \emph {et~al.}(1998)\citenamefont
  {Cohen-Tannoudji}, \citenamefont {Dupont-Roc},\ and\ \citenamefont
  {Grynberg}}]{Cohen-Tannoudji1998}%
  \BibitemOpen
  \bibfield  {author} {\bibinfo {author} {\bibfnamefont {C.}~\bibnamefont
  {Cohen-Tannoudji}}, \bibinfo {author} {\bibfnamefont {J.}~\bibnamefont
  {Dupont-Roc}}, \ and\ \bibinfo {author} {\bibfnamefont {G.}~\bibnamefont
  {Grynberg}},\ }\href {\doibase 10.1002/9783527617197} {\emph {\bibinfo
  {title} {{Atom-Photon Interactions}}}}\ (\bibinfo  {publisher} {Wiley-VCH
  Verlag GmbH},\ \bibinfo {address} {Weinheim, Germany},\ \bibinfo {year}
  {1998})\BibitemShut {NoStop}%
\bibitem [{\citenamefont {Hofmann}\ and\ \citenamefont
  {Takeuchi}(2003)}]{Hofmann2003}%
  \BibitemOpen
  \bibfield  {author} {\bibinfo {author} {\bibfnamefont {H.~F.}\ \bibnamefont
  {Hofmann}}\ and\ \bibinfo {author} {\bibfnamefont {S.}~\bibnamefont
  {Takeuchi}},\ }\bibfield  {title} {\enquote {\bibinfo {title} {{Violation of
  local uncertainty relations as a signature of entanglement}},}\ }\href
  {\doibase 10.1103/PhysRevA.68.032103} {\bibfield  {journal} {\bibinfo
  {journal} {Phys. Rev. A}\ }\textbf {\bibinfo {volume} {68}},\ \bibinfo
  {pages} {032103} (\bibinfo {year} {2003})}\BibitemShut {NoStop}%
\bibitem [{\citenamefont {Dorfman}\ \emph {et~al.}(2016)\citenamefont
  {Dorfman}, \citenamefont {Schlawin},\ and\ \citenamefont
  {Mukamel}}]{Dorfman2016}%
  \BibitemOpen
  \bibfield  {author} {\bibinfo {author} {\bibfnamefont {K.~E.}\ \bibnamefont
  {Dorfman}}, \bibinfo {author} {\bibfnamefont {F.}~\bibnamefont {Schlawin}}, \
  and\ \bibinfo {author} {\bibfnamefont {S.}~\bibnamefont {Mukamel}},\
  }\bibfield  {title} {\enquote {\bibinfo {title} {{Nonlinear optical signals
  and spectroscopy with quantum light}},}\ }\href {\doibase
  10.1103/RevModPhys.88.045008} {\bibfield  {journal} {\bibinfo  {journal}
  {Rev. Mod. Phys.}\ }\textbf {\bibinfo {volume} {88}},\ \bibinfo {pages}
  {045008} (\bibinfo {year} {2016})}\BibitemShut {NoStop}%
\bibitem [{\citenamefont {Chu}\ \emph {et~al.}(2017)\citenamefont {Chu},
  \citenamefont {Götzinger},\ and\ \citenamefont {Sandoghdar}}]{Chu2017}%
  \BibitemOpen
  \bibfield  {author} {\bibinfo {author} {\bibfnamefont {X.-L.}\ \bibnamefont
  {Chu}}, \bibinfo {author} {\bibfnamefont {S.}~\bibnamefont {Götzinger}}, \
  and\ \bibinfo {author} {\bibfnamefont {V.}~\bibnamefont {Sandoghdar}},\
  }\bibfield  {title} {\enquote {\bibinfo {title} {{A single molecule as a
  high-fidelity photon gun for producing intensity-squeezed light}},}\ }\href
  {\doibase 10.1038/nphoton.2016.236} {\bibfield  {journal} {\bibinfo
  {journal} {Nat. Photonics}\ }\textbf {\bibinfo {volume} {11}},\ \bibinfo
  {pages} {58--62} (\bibinfo {year} {2017})}\BibitemShut {NoStop}%
\bibitem [{\citenamefont {Gibbs}\ \emph {et~al.}(1979)\citenamefont {Gibbs},
  \citenamefont {McCall},\ and\ \citenamefont {Venkatesan}}]{Gibbs1979}%
  \BibitemOpen
  \bibfield  {author} {\bibinfo {author} {\bibfnamefont {H.~M.}\ \bibnamefont
  {Gibbs}}, \bibinfo {author} {\bibfnamefont {S.~L.}\ \bibnamefont {McCall}}, \
  and\ \bibinfo {author} {\bibfnamefont {T.~N.~C.}\ \bibnamefont
  {Venkatesan}},\ }\bibfield  {title} {\enquote {\bibinfo {title} {{Optical
  Bistability}},}\ }\href {\doibase 10.1364/ON.5.3.000006} {\bibfield
  {journal} {\bibinfo  {journal} {Optics News}\ }\textbf {\bibinfo {volume}
  {5}},\ \bibinfo {pages} {6} (\bibinfo {year} {1979})}\BibitemShut {NoStop}%
\bibitem [{\citenamefont {Hwang}\ \emph {et~al.}(2009)\citenamefont {Hwang},
  \citenamefont {Pototschnig}, \citenamefont {Lettow}, \citenamefont {Zumofen},
  \citenamefont {Renn}, \citenamefont {Götzinger},\ and\ \citenamefont
  {Sandoghdar}}]{Hwang2009}%
  \BibitemOpen
  \bibfield  {author} {\bibinfo {author} {\bibfnamefont {J.}~\bibnamefont
  {Hwang}}, \bibinfo {author} {\bibfnamefont {M.}~\bibnamefont {Pototschnig}},
  \bibinfo {author} {\bibfnamefont {R.}~\bibnamefont {Lettow}}, \bibinfo
  {author} {\bibfnamefont {G.}~\bibnamefont {Zumofen}}, \bibinfo {author}
  {\bibfnamefont {A.}~\bibnamefont {Renn}}, \bibinfo {author} {\bibfnamefont
  {S.}~\bibnamefont {Götzinger}}, \ and\ \bibinfo {author} {\bibfnamefont
  {V.}~\bibnamefont {Sandoghdar}},\ }\bibfield  {title} {\enquote {\bibinfo
  {title} {{A single-molecule optical transistor}},}\ }\href {\doibase
  10.1038/nature08134} {\bibfield  {journal} {\bibinfo  {journal} {Nature}\
  }\textbf {\bibinfo {volume} {460}},\ \bibinfo {pages} {76--80} (\bibinfo
  {year} {2009})}\BibitemShut {NoStop}%
\bibitem [{\citenamefont {Chuang}\ and\ \citenamefont
  {Yamamoto}(1995)}]{Chuang1995}%
  \BibitemOpen
  \bibfield  {author} {\bibinfo {author} {\bibfnamefont {I.~L.}\ \bibnamefont
  {Chuang}}\ and\ \bibinfo {author} {\bibfnamefont {Y.}~\bibnamefont
  {Yamamoto}},\ }\bibfield  {title} {\enquote {\bibinfo {title} {{Simple
  quantum computer}},}\ }\href {\doibase 10.1103/PhysRevA.52.3489} {\bibfield
  {journal} {\bibinfo  {journal} {Phys. Rev. A}\ }\textbf {\bibinfo {volume}
  {52}},\ \bibinfo {pages} {3489--3496} (\bibinfo {year} {1995})}\BibitemShut
  {NoStop}%
\bibitem [{\citenamefont {Shapiro}(2006)}]{Shapiro2006}%
  \BibitemOpen
  \bibfield  {author} {\bibinfo {author} {\bibfnamefont {J.~H.}\ \bibnamefont
  {Shapiro}},\ }\bibfield  {title} {\enquote {\bibinfo {title} {{Single-photon
  Kerr nonlinearities do not help quantum computation}},}\ }\href {\doibase
  10.1103/PhysRevA.73.062305} {\bibfield  {journal} {\bibinfo  {journal} {Phys.
  Rev. A}\ }\textbf {\bibinfo {volume} {73}},\ \bibinfo {pages} {062305}
  (\bibinfo {year} {2006})}\BibitemShut {NoStop}%
\bibitem [{\citenamefont {Gea-Banacloche}(2010)}]{Gea-Banacloche2010}%
  \BibitemOpen
  \bibfield  {author} {\bibinfo {author} {\bibfnamefont {J.}~\bibnamefont
  {Gea-Banacloche}},\ }\bibfield  {title} {\enquote {\bibinfo {title}
  {{Impossibility of large phase shifts via the giant Kerr effect with
  single-photon wave packets}},}\ }\href {\doibase 10.1103/PhysRevA.81.043823}
  {\bibfield  {journal} {\bibinfo  {journal} {Phys. Rev. A}\ }\textbf {\bibinfo
  {volume} {81}},\ \bibinfo {pages} {043823} (\bibinfo {year}
  {2010})}\BibitemShut {NoStop}%
\bibitem [{\citenamefont {Brod}\ and\ \citenamefont {Combes}(2016)}]{Brod2016}%
  \BibitemOpen
  \bibfield  {author} {\bibinfo {author} {\bibfnamefont {D.~J.}\ \bibnamefont
  {Brod}}\ and\ \bibinfo {author} {\bibfnamefont {J.}~\bibnamefont {Combes}},\
  }\bibfield  {title} {\enquote {\bibinfo {title} {{Passive CPHASE Gate via
  Cross-Kerr Nonlinearities}},}\ }\href {\doibase
  10.1103/PhysRevLett.117.080502} {\bibfield  {journal} {\bibinfo  {journal}
  {Phys. Rev. Lett.}\ }\textbf {\bibinfo {volume} {117}},\ \bibinfo {pages}
  {080502} (\bibinfo {year} {2016})}\BibitemShut {NoStop}%
\bibitem [{\citenamefont {Hardy}\ and\ \citenamefont
  {Shamir}(2007)}]{Hardy2007}%
  \BibitemOpen
  \bibfield  {author} {\bibinfo {author} {\bibfnamefont {J.}~\bibnamefont
  {Hardy}}\ and\ \bibinfo {author} {\bibfnamefont {J.}~\bibnamefont {Shamir}},\
  }\bibfield  {title} {\enquote {\bibinfo {title} {{Optics inspired logic
  architecture}},}\ }\href {\doibase 10.1364/OE.15.000150} {\bibfield
  {journal} {\bibinfo  {journal} {Opt. Express}\ }\textbf {\bibinfo {volume}
  {15}},\ \bibinfo {pages} {150} (\bibinfo {year} {2007})}\BibitemShut
  {NoStop}%
\bibitem [{\citenamefont {Touch}\ \emph {et~al.}(2017)\citenamefont {Touch},
  \citenamefont {Cao}, \citenamefont {Ziyadi}, \citenamefont {Almaiman},
  \citenamefont {Mohajerin-Ariaei},\ and\ \citenamefont {Willner}}]{Touch2017}%
  \BibitemOpen
  \bibfield  {author} {\bibinfo {author} {\bibfnamefont {J.}~\bibnamefont
  {Touch}}, \bibinfo {author} {\bibfnamefont {Y.}~\bibnamefont {Cao}}, \bibinfo
  {author} {\bibfnamefont {M.}~\bibnamefont {Ziyadi}}, \bibinfo {author}
  {\bibfnamefont {A.}~\bibnamefont {Almaiman}}, \bibinfo {author}
  {\bibfnamefont {A.}~\bibnamefont {Mohajerin-Ariaei}}, \ and\ \bibinfo
  {author} {\bibfnamefont {A.~E.}\ \bibnamefont {Willner}},\ }\bibfield
  {title} {\enquote {\bibinfo {title} {{Digital optical processing of optical
  communications: towards an Optical Turing Machine}},}\ }\href {\doibase
  10.1515/nanoph-2016-0145} {\bibfield  {journal} {\bibinfo  {journal}
  {Nanophotonics}\ }\textbf {\bibinfo {volume} {6}},\ \bibinfo {pages}
  {507--530} (\bibinfo {year} {2017})}\BibitemShut {NoStop}%
\bibitem [{\citenamefont {Toninelli}\ \emph {et~al.}(2020)\citenamefont
  {Toninelli}, \citenamefont {Gerhardt}, \citenamefont {Clark}, \citenamefont
  {Reserbat-Plantey}, \citenamefont {Götzinger}, \citenamefont {Ristanovic},
  \citenamefont {Colautti}, \citenamefont {Lombardi}, \citenamefont {Major},
  \citenamefont {Deperasińska}, \citenamefont {Pernice}, \citenamefont
  {Koppens}, \citenamefont {Kozankiewicz}, \citenamefont {Gourdon},
  \citenamefont {Sandoghdar},\ and\ \citenamefont
  {Orrit}}]{Toninelli2020arxiv}%
  \BibitemOpen
  \bibfield  {author} {\bibinfo {author} {\bibfnamefont {C.}~\bibnamefont
  {Toninelli}}, \bibinfo {author} {\bibfnamefont {I.}~\bibnamefont {Gerhardt}},
  \bibinfo {author} {\bibfnamefont {A.~S.}\ \bibnamefont {Clark}}, \bibinfo
  {author} {\bibfnamefont {A.}~\bibnamefont {Reserbat-Plantey}}, \bibinfo
  {author} {\bibfnamefont {S.}~\bibnamefont {Götzinger}}, \bibinfo {author}
  {\bibfnamefont {Z.}~\bibnamefont {Ristanovic}}, \bibinfo {author}
  {\bibfnamefont {M.}~\bibnamefont {Colautti}}, \bibinfo {author}
  {\bibfnamefont {P.}~\bibnamefont {Lombardi}}, \bibinfo {author}
  {\bibfnamefont {K.~D.}\ \bibnamefont {Major}}, \bibinfo {author}
  {\bibfnamefont {I.}~\bibnamefont {Deperasińska}}, \bibinfo {author}
  {\bibfnamefont {W.~H.}\ \bibnamefont {Pernice}}, \bibinfo {author}
  {\bibfnamefont {F.~H.~L.}\ \bibnamefont {Koppens}}, \bibinfo {author}
  {\bibfnamefont {B.}~\bibnamefont {Kozankiewicz}}, \bibinfo {author}
  {\bibfnamefont {A.}~\bibnamefont {Gourdon}}, \bibinfo {author} {\bibfnamefont
  {V.}~\bibnamefont {Sandoghdar}}, \ and\ \bibinfo {author} {\bibfnamefont
  {M.}~\bibnamefont {Orrit}},\ }\bibfield  {title} {\enquote {\bibinfo {title}
  {Single organic molecules for photonic quantum technologies},}\ }\href@noop
  {} {\  (\bibinfo {year} {2020})},\ \Eprint {http://arxiv.org/abs/2011.05059}
  {arXiv:2011.05059} \BibitemShut {NoStop}%
\bibitem [{\citenamefont {Haakh}\ \emph {et~al.}(2016)\citenamefont {Haakh},
  \citenamefont {Faez},\ and\ \citenamefont {Sandoghdar}}]{Haakh2016}%
  \BibitemOpen
  \bibfield  {author} {\bibinfo {author} {\bibfnamefont {H.~R}\ \bibnamefont
  {Haakh}}, \bibinfo {author} {\bibfnamefont {S.}~\bibnamefont {Faez}}, \ and\
  \bibinfo {author} {\bibfnamefont {V.}~\bibnamefont {Sandoghdar}},\ }\bibfield
   {title} {\enquote {\bibinfo {title} {{Polaritonic normal-mode splitting and
  light localization in a one-dimensional nanoguide}},}\ }\href {\doibase
  10.1103/PhysRevA.94.053840} {\bibfield  {journal} {\bibinfo  {journal} {Phys.
  Rev. A}\ }\textbf {\bibinfo {volume} {94}},\ \bibinfo {pages} {053840}
  (\bibinfo {year} {2016})}\BibitemShut {NoStop}%
\bibitem [{\citenamefont {Evans}\ \emph {et~al.}(2018)\citenamefont {Evans},
  \citenamefont {Bhaskar}, \citenamefont {Sukachev}, \citenamefont {Nguyen},
  \citenamefont {Sipahigil}, \citenamefont {Burek}, \citenamefont {Machielse},
  \citenamefont {Zhang}, \citenamefont {Zibrov}, \citenamefont {Bielejec},
  \citenamefont {Park}, \citenamefont {Lončar},\ and\ \citenamefont
  {Lukin}}]{Evans2018}%
  \BibitemOpen
  \bibfield  {author} {\bibinfo {author} {\bibfnamefont {R.~E.}\ \bibnamefont
  {Evans}}, \bibinfo {author} {\bibfnamefont {M.~K.}\ \bibnamefont {Bhaskar}},
  \bibinfo {author} {\bibfnamefont {D.~D.}\ \bibnamefont {Sukachev}}, \bibinfo
  {author} {\bibfnamefont {C.~T.}\ \bibnamefont {Nguyen}}, \bibinfo {author}
  {\bibfnamefont {A.}~\bibnamefont {Sipahigil}}, \bibinfo {author}
  {\bibfnamefont {M.~J.}\ \bibnamefont {Burek}}, \bibinfo {author}
  {\bibfnamefont {B.}~\bibnamefont {Machielse}}, \bibinfo {author}
  {\bibfnamefont {G.~H.}\ \bibnamefont {Zhang}}, \bibinfo {author}
  {\bibfnamefont {A.~S.}\ \bibnamefont {Zibrov}}, \bibinfo {author}
  {\bibfnamefont {E.}~\bibnamefont {Bielejec}}, \bibinfo {author}
  {\bibfnamefont {H.}~\bibnamefont {Park}}, \bibinfo {author} {\bibfnamefont
  {M.}~\bibnamefont {Lončar}}, \ and\ \bibinfo {author} {\bibfnamefont
  {M.~D.}\ \bibnamefont {Lukin}},\ }\bibfield  {title} {\enquote {\bibinfo
  {title} {{Photon-mediated interactions between quantum emitters in a diamond
  nanocavity}},}\ }\href {\doibase 10.1126/science.aau4691} {\bibfield
  {journal} {\bibinfo  {journal} {Science}\ }\textbf {\bibinfo {volume}
  {362}},\ \bibinfo {pages} {662--665} (\bibinfo {year} {2018})}\BibitemShut
  {NoStop}%
\bibitem [{\citenamefont {Samutpraphoot}\ \emph {et~al.}(2020)\citenamefont
  {Samutpraphoot}, \citenamefont {Đorđević}, \citenamefont {Ocola},
  \citenamefont {Bernien}, \citenamefont {Senko}, \citenamefont {Vuletić},\
  and\ \citenamefont {Lukin}}]{Samutpraphoot2020}%
  \BibitemOpen
  \bibfield  {author} {\bibinfo {author} {\bibfnamefont {P.}~\bibnamefont
  {Samutpraphoot}}, \bibinfo {author} {\bibfnamefont {T.}~\bibnamefont
  {Đorđević}}, \bibinfo {author} {\bibfnamefont {P.~L.}\ \bibnamefont
  {Ocola}}, \bibinfo {author} {\bibfnamefont {H.}~\bibnamefont {Bernien}},
  \bibinfo {author} {\bibfnamefont {C.}~\bibnamefont {Senko}}, \bibinfo
  {author} {\bibfnamefont {V.}~\bibnamefont {Vuletić}}, \ and\ \bibinfo
  {author} {\bibfnamefont {M.~D.}\ \bibnamefont {Lukin}},\ }\bibfield  {title}
  {\enquote {\bibinfo {title} {{Strong Coupling of Two Individually Controlled
  Atoms via a Nanophotonic Cavity}},}\ }\href {\doibase
  10.1103/PhysRevLett.124.063602} {\bibfield  {journal} {\bibinfo  {journal}
  {Phys. Rev. Lett.}\ }\textbf {\bibinfo {volume} {124}},\ \bibinfo {pages}
  {063602} (\bibinfo {year} {2020})}\BibitemShut {NoStop}%
\bibitem [{\citenamefont {Carusotto}\ and\ \citenamefont
  {Ciuti}(2013)}]{Carusotto2013}%
  \BibitemOpen
  \bibfield  {author} {\bibinfo {author} {\bibfnamefont {I.}~\bibnamefont
  {Carusotto}}\ and\ \bibinfo {author} {\bibfnamefont {C.}~\bibnamefont
  {Ciuti}},\ }\bibfield  {title} {\enquote {\bibinfo {title} {{Quantum fluids
  of light}},}\ }\href {\doibase 10.1103/RevModPhys.85.299} {\bibfield
  {journal} {\bibinfo  {journal} {Rev. Mod. Phys.}\ }\textbf {\bibinfo {volume}
  {85}},\ \bibinfo {pages} {299--366} (\bibinfo {year} {2013})}\BibitemShut
  {NoStop}%
\bibitem [{\citenamefont {Muñoz}\ \emph {et~al.}(2014)\citenamefont {Muñoz},
  \citenamefont {del Valle}, \citenamefont {Tudela}, \citenamefont {Müller},
  \citenamefont {Lichtmannecker}, \citenamefont {Kaniber}, \citenamefont
  {Tejedor}, \citenamefont {Finley},\ and\ \citenamefont {Laussy}}]{Munoz2014}%
  \BibitemOpen
  \bibfield  {author} {\bibinfo {author} {\bibfnamefont {C.~S.}\ \bibnamefont
  {Muñoz}}, \bibinfo {author} {\bibfnamefont {E.}~\bibnamefont {del Valle}},
  \bibinfo {author} {\bibfnamefont {A.~G.}\ \bibnamefont {Tudela}}, \bibinfo
  {author} {\bibfnamefont {K.}~\bibnamefont {Müller}}, \bibinfo {author}
  {\bibfnamefont {S.}~\bibnamefont {Lichtmannecker}}, \bibinfo {author}
  {\bibfnamefont {M.}~\bibnamefont {Kaniber}}, \bibinfo {author} {\bibfnamefont
  {C.}~\bibnamefont {Tejedor}}, \bibinfo {author} {\bibfnamefont {J.~J.}\
  \bibnamefont {Finley}}, \ and\ \bibinfo {author} {\bibfnamefont {F.~P.}\
  \bibnamefont {Laussy}},\ }\bibfield  {title} {\enquote {\bibinfo {title}
  {{Emitters of N-photon bundles}},}\ }\href {\doibase
  10.1038/nphoton.2014.114} {\bibfield  {journal} {\bibinfo  {journal} {Nat.
  Photonics}\ }\textbf {\bibinfo {volume} {8}},\ \bibinfo {pages} {550--555}
  (\bibinfo {year} {2014})}\BibitemShut {NoStop}%
\bibitem [{\citenamefont {Rattenbacher}\ \emph {et~al.}(2019)\citenamefont
  {Rattenbacher}, \citenamefont {Shkarin}, \citenamefont {Renger},
  \citenamefont {Utikal}, \citenamefont {Götzinger},\ and\ \citenamefont
  {Sandoghdar}}]{Rattenbacher2019}%
  \BibitemOpen
  \bibfield  {author} {\bibinfo {author} {\bibfnamefont {D.}~\bibnamefont
  {Rattenbacher}}, \bibinfo {author} {\bibfnamefont {A.}~\bibnamefont
  {Shkarin}}, \bibinfo {author} {\bibfnamefont {J.}~\bibnamefont {Renger}},
  \bibinfo {author} {\bibfnamefont {T.}~\bibnamefont {Utikal}}, \bibinfo
  {author} {\bibfnamefont {S.}~\bibnamefont {Götzinger}}, \ and\ \bibinfo
  {author} {\bibfnamefont {V.}~\bibnamefont {Sandoghdar}},\ }\bibfield  {title}
  {\enquote {\bibinfo {title} {{Coherent coupling of single molecules to
  on-chip ring resonators}},}\ }\href {\doibase 10.1088/1367-2630/ab28b2}
  {\bibfield  {journal} {\bibinfo  {journal} {New J. Phys.}\ }\textbf {\bibinfo
  {volume} {21}},\ \bibinfo {pages} {062002} (\bibinfo {year}
  {2019})}\BibitemShut {NoStop}%
\bibitem [{\citenamefont {Shkarin}\ \emph {et~al.}(2021)\citenamefont
  {Shkarin}, \citenamefont {Rattenbacher}, \citenamefont {Renger},
  \citenamefont {Hönl}, \citenamefont {Utikal}, \citenamefont {Seidler},
  \citenamefont {Götzinger},\ and\ \citenamefont {Sandoghdar}}]{Shkarin2021}%
  \BibitemOpen
  \bibfield  {author} {\bibinfo {author} {\bibfnamefont {A.}~\bibnamefont
  {Shkarin}}, \bibinfo {author} {\bibfnamefont {D.}~\bibnamefont
  {Rattenbacher}}, \bibinfo {author} {\bibfnamefont {J.}~\bibnamefont
  {Renger}}, \bibinfo {author} {\bibfnamefont {S.}~\bibnamefont {Hönl}},
  \bibinfo {author} {\bibfnamefont {T.}~\bibnamefont {Utikal}}, \bibinfo
  {author} {\bibfnamefont {P.}~\bibnamefont {Seidler}}, \bibinfo {author}
  {\bibfnamefont {S.}~\bibnamefont {Götzinger}}, \ and\ \bibinfo {author}
  {\bibfnamefont {V.}~\bibnamefont {Sandoghdar}},\ }\bibfield  {title}
  {\enquote {\bibinfo {title} {{Nanoscopic Charge Fluctuations in a Gallium
  Phosphide Waveguide Measured by Single Molecules}},}\ }\href {\doibase
  10.1103/PhysRevLett.126.133602} {\bibfield  {journal} {\bibinfo  {journal}
  {Physical Review Letters}\ }\textbf {\bibinfo {volume} {126}},\ \bibinfo
  {pages} {133602} (\bibinfo {year} {2021})}\BibitemShut {NoStop}%
\bibitem [{\citenamefont {Minzioni}\ \emph {et~al.}(2019)\citenamefont
  {Minzioni}, \citenamefont {Lacava}, \citenamefont {Tanabe}, \citenamefont
  {Dong}, \citenamefont {Hu}, \citenamefont {Csaba}, \citenamefont {Porod},
  \citenamefont {Singh}, \citenamefont {Willner}, \citenamefont {Almaiman},
  \citenamefont {Torres-Company}, \citenamefont {Schröder}, \citenamefont
  {Peacock}, \citenamefont {Strain}, \citenamefont {Parmigiani}, \citenamefont
  {Contestabile}, \citenamefont {Marpaung}, \citenamefont {Liu}, \citenamefont
  {Bowers}, \citenamefont {Chang}, \citenamefont {Fabbri}, \citenamefont
  {{Ramos Vázquez}}, \citenamefont {Bharadwaj}, \citenamefont {Eaton},
  \citenamefont {Lodahl}, \citenamefont {Zhang}, \citenamefont {Eggleton},
  \citenamefont {Munro}, \citenamefont {Nemoto}, \citenamefont {Morin},
  \citenamefont {Laurat},\ and\ \citenamefont {Nunn}}]{Minzioni2019}%
  \BibitemOpen
  \bibfield  {author} {\bibinfo {author} {\bibfnamefont {P.}~\bibnamefont
  {Minzioni}}, \bibinfo {author} {\bibfnamefont {C.}~\bibnamefont {Lacava}},
  \bibinfo {author} {\bibfnamefont {T.}~\bibnamefont {Tanabe}}, \bibinfo
  {author} {\bibfnamefont {J.}~\bibnamefont {Dong}}, \bibinfo {author}
  {\bibfnamefont {X.}~\bibnamefont {Hu}}, \bibinfo {author} {\bibfnamefont
  {G.}~\bibnamefont {Csaba}}, \bibinfo {author} {\bibfnamefont
  {W.}~\bibnamefont {Porod}}, \bibinfo {author} {\bibfnamefont
  {G.}~\bibnamefont {Singh}}, \bibinfo {author} {\bibfnamefont {A.~E.}\
  \bibnamefont {Willner}}, \bibinfo {author} {\bibfnamefont {A.}~\bibnamefont
  {Almaiman}}, \bibinfo {author} {\bibfnamefont {V.}~\bibnamefont
  {Torres-Company}}, \bibinfo {author} {\bibfnamefont {J.}~\bibnamefont
  {Schröder}}, \bibinfo {author} {\bibfnamefont {A.~C.}\ \bibnamefont
  {Peacock}}, \bibinfo {author} {\bibfnamefont {M.~J.}\ \bibnamefont {Strain}},
  \bibinfo {author} {\bibfnamefont {F.}~\bibnamefont {Parmigiani}}, \bibinfo
  {author} {\bibfnamefont {G.}~\bibnamefont {Contestabile}}, \bibinfo {author}
  {\bibfnamefont {D.}~\bibnamefont {Marpaung}}, \bibinfo {author}
  {\bibfnamefont {Z.}~\bibnamefont {Liu}}, \bibinfo {author} {\bibfnamefont
  {J.~E.}\ \bibnamefont {Bowers}}, \bibinfo {author} {\bibfnamefont
  {L.}~\bibnamefont {Chang}}, \bibinfo {author} {\bibfnamefont
  {S.}~\bibnamefont {Fabbri}}, \bibinfo {author} {\bibfnamefont
  {M.}~\bibnamefont {{Ramos Vázquez}}}, \bibinfo {author} {\bibfnamefont
  {V.}~\bibnamefont {Bharadwaj}}, \bibinfo {author} {\bibfnamefont {S.~M.}\
  \bibnamefont {Eaton}}, \bibinfo {author} {\bibfnamefont {P.}~\bibnamefont
  {Lodahl}}, \bibinfo {author} {\bibfnamefont {X.}~\bibnamefont {Zhang}},
  \bibinfo {author} {\bibfnamefont {B.~J.}\ \bibnamefont {Eggleton}}, \bibinfo
  {author} {\bibfnamefont {W.~J.}\ \bibnamefont {Munro}}, \bibinfo {author}
  {\bibfnamefont {K.}~\bibnamefont {Nemoto}}, \bibinfo {author} {\bibfnamefont
  {O.}~\bibnamefont {Morin}}, \bibinfo {author} {\bibfnamefont
  {J.}~\bibnamefont {Laurat}}, \ and\ \bibinfo {author} {\bibfnamefont
  {J.}~\bibnamefont {Nunn}},\ }\bibfield  {title} {\enquote {\bibinfo {title}
  {{Roadmap on all-optical processing}},}\ }\href {\doibase
  10.1088/2040-8986/ab0e66} {\bibfield  {journal} {\bibinfo  {journal} {J.
  Opt.}\ }\textbf {\bibinfo {volume} {21}},\ \bibinfo {pages} {063001}
  (\bibinfo {year} {2019})}\BibitemShut {NoStop}%
\bibitem [{\citenamefont {Savage}\ and\ \citenamefont
  {Carmichael}(1988)}]{SavageCarmichael1988}%
  \BibitemOpen
  \bibfield  {author} {\bibinfo {author} {\bibfnamefont {C.~M.}\ \bibnamefont
  {Savage}}\ and\ \bibinfo {author} {\bibfnamefont {H.~J.}\ \bibnamefont
  {Carmichael}},\ }\bibfield  {title} {\enquote {\bibinfo {title} {{Single atom
  optical bistability}},}\ }\href {\doibase 10.1109/3.7075} {\bibfield
  {journal} {\bibinfo  {journal} {IEEE J. Quantum Electron.}\ }\textbf
  {\bibinfo {volume} {24}},\ \bibinfo {pages} {1495--1498} (\bibinfo {year}
  {1988})}\BibitemShut {NoStop}%
\bibitem [{\citenamefont {Rice}\ and\ \citenamefont
  {Carmichael}(1988)}]{Rice1988}%
  \BibitemOpen
  \bibfield  {author} {\bibinfo {author} {\bibfnamefont {P.~R.}\ \bibnamefont
  {Rice}}\ and\ \bibinfo {author} {\bibfnamefont {H.~J.}\ \bibnamefont
  {Carmichael}},\ }\bibfield  {title} {\enquote {\bibinfo {title} {{Single-atom
  cavity-enhanced absorption. I. Photon statistics in the bad-cavity limit}},}\
  }\href {\doibase 10.1109/3.974} {\bibfield  {journal} {\bibinfo  {journal}
  {IEEE J. Quantum Electron.}\ }\textbf {\bibinfo {volume} {24}},\ \bibinfo
  {pages} {1351--1366} (\bibinfo {year} {1988})}\BibitemShut {NoStop}%
\bibitem [{\citenamefont {Lindberg}\ and\ \citenamefont
  {Savage}(1988)}]{Lindberg1988}%
  \BibitemOpen
  \bibfield  {author} {\bibinfo {author} {\bibfnamefont {M.}~\bibnamefont
  {Lindberg}}\ and\ \bibinfo {author} {\bibfnamefont {C.~M.}\ \bibnamefont
  {Savage}},\ }\bibfield  {title} {\enquote {\bibinfo {title} {{Steady-state
  two-level atomic population inversion via a quantized cavity field}},}\
  }\href {\doibase 10.1103/PhysRevA.38.5182} {\bibfield  {journal} {\bibinfo
  {journal} {Phys. Rev. A}\ }\textbf {\bibinfo {volume} {38}},\ \bibinfo
  {pages} {5182--5192} (\bibinfo {year} {1988})}\BibitemShut {NoStop}%
\bibitem [{\citenamefont {Mu}\ and\ \citenamefont {Savage}(1992)}]{Mu1992}%
  \BibitemOpen
  \bibfield  {author} {\bibinfo {author} {\bibfnamefont {Y.}~\bibnamefont
  {Mu}}\ and\ \bibinfo {author} {\bibfnamefont {C.~M.}\ \bibnamefont
  {Savage}},\ }\bibfield  {title} {\enquote {\bibinfo {title} {{One-atom
  lasers}},}\ }\href {\doibase 10.1103/PhysRevA.46.5944} {\bibfield  {journal}
  {\bibinfo  {journal} {Phys. Rev. A}\ }\textbf {\bibinfo {volume} {46}},\
  \bibinfo {pages} {5944--5954} (\bibinfo {year} {1992})}\BibitemShut {NoStop}%
\bibitem [{\citenamefont {An}\ \emph {et~al.}(1994)\citenamefont {An},
  \citenamefont {Childs}, \citenamefont {Dasari},\ and\ \citenamefont
  {Feld}}]{An1994}%
  \BibitemOpen
  \bibfield  {author} {\bibinfo {author} {\bibfnamefont {K.}~\bibnamefont
  {An}}, \bibinfo {author} {\bibfnamefont {J.~J.}\ \bibnamefont {Childs}},
  \bibinfo {author} {\bibfnamefont {R.~R.}\ \bibnamefont {Dasari}}, \ and\
  \bibinfo {author} {\bibfnamefont {M.~S.}\ \bibnamefont {Feld}},\ }\bibfield
  {title} {\enquote {\bibinfo {title} {{Microlaser: A laser with One Atom in an
  Optical Resonator}},}\ }\href {\doibase 10.1103/PhysRevLett.73.3375}
  {\bibfield  {journal} {\bibinfo  {journal} {Phys. Rev. Lett.}\ }\textbf
  {\bibinfo {volume} {73}},\ \bibinfo {pages} {3375--3378} (\bibinfo {year}
  {1994})}\BibitemShut {NoStop}%
\bibitem [{\citenamefont {Nomura}\ \emph {et~al.}(2010)\citenamefont {Nomura},
  \citenamefont {Kumagai}, \citenamefont {Iwamoto}, \citenamefont {Ota},\ and\
  \citenamefont {Arakawa}}]{Nomura2010}%
  \BibitemOpen
  \bibfield  {author} {\bibinfo {author} {\bibfnamefont {M.}~\bibnamefont
  {Nomura}}, \bibinfo {author} {\bibfnamefont {N.}~\bibnamefont {Kumagai}},
  \bibinfo {author} {\bibfnamefont {S.}~\bibnamefont {Iwamoto}}, \bibinfo
  {author} {\bibfnamefont {Y.}~\bibnamefont {Ota}}, \ and\ \bibinfo {author}
  {\bibfnamefont {Y.}~\bibnamefont {Arakawa}},\ }\bibfield  {title} {\enquote
  {\bibinfo {title} {{Laser oscillation in a strongly coupled
  single-quantum-dot–nanocavity system}},}\ }\href {\doibase
  10.1038/nphys1518} {\bibfield  {journal} {\bibinfo  {journal} {Nat. Phys.}\
  }\textbf {\bibinfo {volume} {6}},\ \bibinfo {pages} {279--283} (\bibinfo
  {year} {2010})}\BibitemShut {NoStop}%
\bibitem [{\citenamefont {McKeever}\ \emph {et~al.}(2003)\citenamefont
  {McKeever}, \citenamefont {Boca}, \citenamefont {Boozer}, \citenamefont
  {Buck},\ and\ \citenamefont {Kimble}}]{McKeever2003}%
  \BibitemOpen
  \bibfield  {author} {\bibinfo {author} {\bibfnamefont {J.}~\bibnamefont
  {McKeever}}, \bibinfo {author} {\bibfnamefont {A.}~\bibnamefont {Boca}},
  \bibinfo {author} {\bibfnamefont {A.~D.}\ \bibnamefont {Boozer}}, \bibinfo
  {author} {\bibfnamefont {J.~R.}\ \bibnamefont {Buck}}, \ and\ \bibinfo
  {author} {\bibfnamefont {H.~J.}\ \bibnamefont {Kimble}},\ }\bibfield  {title}
  {\enquote {\bibinfo {title} {{Experimental realization of a one-atom laser in
  the regime of strong coupling}},}\ }\href {\doibase 10.1038/nature01974}
  {\bibfield  {journal} {\bibinfo  {journal} {Nature}\ }\textbf {\bibinfo
  {volume} {425}},\ \bibinfo {pages} {268--271} (\bibinfo {year}
  {2003})}\BibitemShut {NoStop}%
\bibitem [{\citenamefont {Zirkelbach}\ \emph {et~al.}(2020)\citenamefont
  {Zirkelbach}, \citenamefont {Gmeiner}, \citenamefont {Renger}, \citenamefont
  {Türschmann}, \citenamefont {Utikal}, \citenamefont {Götzinger},\ and\
  \citenamefont {Sandoghdar}}]{Zirkelbach2020}%
  \BibitemOpen
  \bibfield  {author} {\bibinfo {author} {\bibfnamefont {J.}~\bibnamefont
  {Zirkelbach}}, \bibinfo {author} {\bibfnamefont {B.}~\bibnamefont {Gmeiner}},
  \bibinfo {author} {\bibfnamefont {J.}~\bibnamefont {Renger}}, \bibinfo
  {author} {\bibfnamefont {P.}~\bibnamefont {Türschmann}}, \bibinfo {author}
  {\bibfnamefont {T.}~\bibnamefont {Utikal}}, \bibinfo {author} {\bibfnamefont
  {S.}~\bibnamefont {Götzinger}}, \ and\ \bibinfo {author} {\bibfnamefont
  {V.}~\bibnamefont {Sandoghdar}},\ }\bibfield  {title} {\enquote {\bibinfo
  {title} {{Partial Cloaking of a Gold Particle by a Single Molecule}},}\
  }\href {\doibase 10.1103/PhysRevLett.125.103603} {\bibfield  {journal}
  {\bibinfo  {journal} {Phys. Rev. Lett.}\ }\textbf {\bibinfo {volume} {125}},\
  \bibinfo {pages} {103603} (\bibinfo {year} {2020})}\BibitemShut {NoStop}%
\bibitem [{\citenamefont {Gürlek}\ \emph {et~al.}(2018)\citenamefont
  {Gürlek}, \citenamefont {Sandoghdar},\ and\ \citenamefont
  {Martín-Cano}}]{Gurlek2018}%
  \BibitemOpen
  \bibfield  {author} {\bibinfo {author} {\bibfnamefont {B.}~\bibnamefont
  {Gürlek}}, \bibinfo {author} {\bibfnamefont {V.}~\bibnamefont {Sandoghdar}},
  \ and\ \bibinfo {author} {\bibfnamefont {D.}~\bibnamefont {Martín-Cano}},\
  }\bibfield  {title} {\enquote {\bibinfo {title} {{Manipulation of Quenching
  in Nanoantenna–Emitter Systems Enabled by External Detuned Cavities: A Path
  to Enhance Strong-Coupling}},}\ }\href {\doibase
  10.1021/acsphotonics.7b00953} {\bibfield  {journal} {\bibinfo  {journal} {ACS
  Photonics}\ }\textbf {\bibinfo {volume} {5}},\ \bibinfo {pages} {456--461}
  (\bibinfo {year} {2018})}\BibitemShut {NoStop}%
\end{thebibliography}

\end{document}